\documentclass{ieeeaccess}
\usepackage{colortbl}
\usepackage{tikz}
\usepackage{pgfplots}
\pgfplotsset{compat=1.18}
\usepackage{tcolorbox}
\usepackage{amsmath,amssymb,amsfonts}
\usepackage{algorithmic}
\usepackage{algorithm}
\usepackage{array}
\usepackage[caption=false,font=normalsize,labelfont=sf,textfont=sf]{subfig}
\usepackage{textcomp}
\usepackage{stfloats}
\usepackage{hyperref}
\usepackage{xurl}
\hypersetup{breaklinks=true}
\usepackage{verbatim}
\usepackage{booktabs}
\usepackage{graphicx}
\usepackage{multirow}
\usepackage{tabularx}
  \newcolumntype{Y}{>{\centering\arraybackslash}X}
\usepackage{makecell}
\usepackage{standalone}
\usepackage{enumitem}
\usepackage{listings}
\usepackage{microtype}
\usepackage[numbers,sort&compress]{natbib}
\usepackage{multibib}
\usepackage{multicol}
\usetikzlibrary{positioning, arrows, arrows.meta, calc, automata, shapes, shadows, fit, backgrounds}
\usepackage[T1]{fontenc}
\usepackage[utf8]{inputenc}

\newcites{r}{Reviewed Literature}

\DeclareFontShape{T1}{formata}{m}{sl} { <-> ssub * formata/m/it }{}
\NewSpotColorSpace{PANTONE}
\AddSpotColor{PANTONE} {PANTONE3015C} {PANTONE\SpotSpace 3015\SpotSpace C} {1 0.3 0 0.2}
\SetPageColorSpace{PANTONE}

\makeatletter
\AtBeginDocument{\DeclareMathVersion{bold}
\SetSymbolFont{operators}{bold}{T1}{times}{b}{n}
\SetSymbolFont{NewLetters}{bold}{T1}{times}{b}{it}
\SetMathAlphabet{\mathrm}{bold}{T1}{times}{b}{n}
\SetMathAlphabet{\mathit}{bold}{T1}{times}{b}{it}
\SetMathAlphabet{\mathbf}{bold}{T1}{times}{b}{n}
\SetMathAlphabet{\mathtt}{bold}{OT1}{pcr}{b}{n}
\SetSymbolFont{symbols}{bold}{OMS}{cmsy}{b}{n}
\renewcommand\boldmath{\@nomath\boldmath\mathversion{bold}}}
\makeatother

\definecolor{blue}{RGB}{5, 107, 138}
\colorlet{blue}{blue!80}
\definecolor{red}{RGB}{187, 68, 48}
\definecolor{green}{RGB}{246, 174, 45}

\newif\ifanonymous
\anonymousfalse
\renewcommand{\thevol}{X}
\renewcommand{\theyear}{X}

\begin{document}

\history{Date of current version 03 November 2025.}
\doi{}

\title{Toward a Common Understanding of Cryptographic Agility --- A Systematic Review}

\ifanonymous
  \author{Anonymous Author}
  \address{Anonymous Institution}
\else
\author{
  \uppercase{Christian Näther}\authorrefmark{1}, 
  \uppercase{Daniel Herzinger}\authorrefmark{2}, 
  \uppercase{Jan-Philipp Steghöfer}\authorrefmark{1}, \IEEEmembership{Member, IEEE},
  \uppercase{Stefan-Lukas Gazdag}\authorrefmark{2}, 
  \uppercase{Eduard Hirsch}\authorrefmark{3}, 
  AND \uppercase{Daniel Loebenberger}\authorrefmark{4}
}

\address[1]{XITASO GmbH, Augsburg, 86153 Germany}
\address[2]{genua GmbH, Kirchheim, 85551 Germany}
\address[3]{Technical University of Applied Sciences Amberg-Weiden, Amberg, 92224 Germany}
\address[4]{Fraunhofer AISEC, Garching, 85748 Germany}

\markboth
{Näther \headeretal: Toward a Common Understanding of Cryptographic Agility -- A Systematic Review}
{Näther \headeretal: Toward a Common Understanding of Cryptographic Agility -- A Systematic Review}

\corresp{Corresponding author: Christian Näther (e-mail: christian.naether@xitaso.com)}

\tfootnote{The work was funded by the German Federal Ministry of Education and Research.}
\fi

\begin{abstract}
Cryptographic agility has become a recurring theme in both academic and industrial discussions, driven by the need to adapt cryptography under evolving threats and disruptive changes such as post-quantum cryptography. 
Yet despite its rising importance, the concept itself remains inconsistently defined and often conflated with related notions, limiting its value for research and practice.

This paper offers the first systematic study of cryptographic agility by providing the following research contributions:
First, we review existing definitions and classify their shared characteristics across 84 screened sources, of which 48 are included in our analysis (24 peer-reviewed, 24 gray literature), thereby exposing the lack of consensus within
the current literature~(RQ1). 
Second, we synthesize these findings into a literature-based definition that unifies the most significant aspects identified in the literature (RQ2). 
Third, we refine this foundation into a general, context-independent definition that isolates agility as the property of changeability of cryptographic entities (RQ3). 
Fourth, we extend the concept by introducing a layer model with four overarching
domains---conceptual, software, hardware, and organization---and by clarifying its relation to neighboring concepts such as cryptographic versatility and interoperability (RQ4).

We demonstrate the applicability of our results by analyzing the three
real-world exemplars OpenSSL, NGINX, and GitLab CI/CD\@.
Finally, we discuss broader implications, including trade-offs with complexity, its assessment across contexts, and the role of agility in cryptographic migration. 
Our work establishes a unified terminology and conceptual foundation for cryptographic agility, enabling consistent reasoning and guiding both research and practice in designing systems that can adapt securely and sustainably.
\end{abstract}
\vspace{.5cm}


\titlepgskip=-21pt
\maketitle

% Save footnote
\newcounter{savefootnote}
\setcounter{savefootnote}{\value{footnote}}

% Create an unenumerated footnote
\renewcommand{\thefootnote}{\relax}%
\footnotetext{This work has been submitted to the IEEE for possible publication. Copyright may be transferred without notice, after which this version may no longer be accessible.}%

% Restore footnote numbering
\setcounter{footnote}{\value{savefootnote}}
\renewcommand{\thefootnote}{\arabic{footnote}}

% We need the following command for the corret order of our reviewed literature
\nociter{Grote19ReviewOfPQC, Richter21AgileAndVersatileQuantum, Badertscher22ComposableApproach, Badari21OverviewBitcoin, Ma21CARAF, Moustafa18CAMustHave, rfc6421, Nationalacademies17CAandInteroperability, MacaulayCAinPractice, ETSI20CYBERMigrationStrategies, Sikeridis23Enterprise-levelCA, Barker21GettingReadyForPQC, rfc7696GuideslinesCA, Ott19ResearchChallenges, Fan21ImpactOfPQCHybridCerts, Vasic16LightweightSolution, Zhang23MakingSWQuantumsafe, ISARA20ManagingCryptoRisk, Wiesmaier21PQCMigration, Paul19ImportanceOfCA, Alnahawi23StateOfCA, Paul21TransitionToPQCIoT, Würth21PQC, Cunningham21SystemAcquisition, digicert20PQCMaturityModel, Heftrig22PosterDNSSEC, Vasic12SecurityAgility, Sullivan09CA, LaMacchia21LongRoadToPQC, Mashatan21ComplexPath, Mehrez18CAProperties, handbook, Hohm23MaturityModelCA, Yunakovsky21TowardsSecurityRecommendations, Heid23TracingCA, Henry18CA, Ott23ResearchOnCryptographicTransition, marchesi2025CAinLightOfQuantumResistenec, cho2024SWDefinedCryptographyDesignFeature, Silonosov2024CAManifestoinE2EE, Alqabandi2024CAinMobileBanking, ras2025CAinHWPhoenix, frauenschlaeger2024CAinOperationalTech, fries2024CAwithAttributeCerts, galambosHWsupportedCAinQKD, sanon2025QuantumReadyMobileCommunications, kourtis2025AdaptivePQCforBlockchain, barker2025NISTConsiderationsforCA}

\section{Introduction}
\label{sec:introduction}
Cryptography is an integral part of our digital world.
It ensures information security across diverse domains, from encryption protocols securing web traffic and cloud storage authentication~\cite{Stallings2022CryptoAndNetworkSec} to technologies such as blockchain~\cite{Zhang20Blockchain}.
Almost every guarantee of trust in the IT landscape relies on cryptography~\cite{kessler2019overview}.

In the near future, the landscape of cryptography is about to change with increasing pace.
Advances in quantum computing~\cite{Wilhelm23StatusQuantumComputing} have the potential to weaken or even break essential cryptographic algorithms by solving the underlying mathematical problems significantly faster than classical computers~\cite{Chen16ReportPQC,Ehlen21KryptografieSicher}. 
The main response is the migration to post-quantum cryptography (PQC)~\cite{bernstein2017post,niederhagen2017practical}.
The urgency of this transition made PQC migration a role model of applied cryptography.
Beyond the PQC migration, attack vectors such as side-channel attacks~\cite{Randolph20SideChannel} and advancements in cryptanalysis~\cite{Dooley18HistoryCryptanalysis} continually challenge the security of established cryptographic systems.
The deprecation of TLS~1.0 and TLS~1.1 provides a concrete illustration: although both versions were widely recognized as insecure, their removal required years of coordinated effort across widely deployed systems before it could be enforced~\cite{rfc8996}.
Governmental agencies and institutions have emphasized the importance of cryptographic agility for several years and consider it a significant aspect for future recommendations and requirements~\cite{Ehlen21KryptografieSicher,white_house_memorandum,bsi_crypto,bsi_handlung,lilychenagility}.

Historical transitions also illustrate why cryptographic agility is essential,
demonstrating both progress and recurring friction.
For example, the move from DSA to ECDSA~\cite{fips186,fips1865} reduced public key sizes~\cite{johnson1998elliptic}, improving applicability in resource-constrained environments. 
Yet even this migration came with challenges~\cite{cryptotransitionlater,dnssec_ecdsa}. 
Recurring issues include web servers interacting with heterogeneous clients~\cite{sowa2024postquantumcryptographypqcnetwork} or compliance with differing national standards such as US~FIPS~\cite{nistFIPSpublications} versus Russian~GOST~\cite{arutyunov2017clustering}.
Protocols negotiating cryptographic algorithms at runtime provide one way to mitigate
such problems, but their implementation and testing can be resource-intensive~\citer{rfc6421}. 
Still, these mechanisms are present in software and standards, suggesting that aspects of cryptographic agility have been addressed, although imperfectly~\cite{cryptotransitionlater,cryptotransitionearly}.

Real-world developments reinforce this perspective. Early PQC migration steps integrated lightweight schemes such as NTRU~Prime~\cite{ntru_prime_nist_submission} in tools like TinySSH\footnote{\url{https://tinyssh.org/}} or OpenSSH\footnote{\url{https://www.openssh.com/}}. 
Broader experiments focused on minimalistic changes (see, e.g.,~\cite{google_tls,cloudflare_tls}). 
Attempts at more agile solutions revealed how complexity increases with agility. 
For example, the current approaches~\cite{rfc9242,rfc9370,beyond_64k,nir-ipsecme-big-payload-04} for making the key agreement protocol
IKEv2~\cite{rfc7296} quantum-safe, apart from using pre-shared keys~\cite{rfc8784},
introduce significant additional logic, which increases protocol complexity,
complicates implementations, and multiplies testing efforts~\cite{ikev2_complexity}. 
Taken together, these examples highlight that without systematic agility, every cryptographic transition risks high cost, limited interoperability, and delayed deployment.

It is therefore not surprising that, in the face of the PQC transition, the number of papers mentioning cryptographic agility is rising. 
Across the literature there is broad consensus that agility is important, if not
a cornerstone of any cryptographic transition strategy.
Yet despite agreement on its importance, the concept itself remains weakly defined.
Most authors pick ad-hoc definitions that suit their individual needs.
For example, some authors treat cryptographic agility as an inherent property of systems, others as an engineering practice, and still others as a broader design paradigm. 
Furthermore, concepts related to cryptographic agility, such as ``cryptographic versatility'' and ``cryptographic interoperability'', are mixed together. 
As a result, inconsistencies arise in both academic research and practical implementation. 
This conceptual ambiguity motivates the central question raised by Ott et al.: 
``Where is the research on cryptographic transition and agility?''~\citer{Ott19ResearchChallenges,Ott23ResearchOnCryptographicTransition}.

To address this issue, we conducted the first systematic study of cryptographic agility. 
Simply put, this paper explains what cryptographic agility is, what it is not, and where it manifests in real systems.
Therefore, we synthesized prior attempts at definition, refined them into a literature-based and subsequently a context-independent definition, clarified their relation to neighboring concepts, operationalized the concept through a layer model, and demonstrated its applicability through three real-world examples. 
Accordingly, our contributions are:
\begin{enumerate}
    \item We illustrate the variety of existing definitions in the literature and derive their fundamental characteristics (cf.~Section~\ref{sec:results:variety_of_definitions}). This highlights the lack of coherence in the literature and the potential for misunderstandings.

    \item We synthesize these findings into a literature-based definition and further
			refine it to a context-independent definition
			(cf.~Sections~\ref{sec:results:context-independent_def_of_ca} and~\ref{sec:results:context-independent_def_ca}).
			This provides a precise basis for future research on cryptographic agility.

    \item We operationalize the concept through a layer model and four overarching domains and position related notions with clear boundaries (cf.~Section~\ref{sec:results:rq3_relationships_of_cryptographic_agility}). This clarifies terminology and enables consistent reasoning across contexts.

    \item We demonstrate the applicability of our results by analyzing three real-world exemplars, namely OpenSSL, NGINX, and GitLab CI/CD (Continuous Integration/Continuous Delivery), within the software domain (cf.~Section~\ref{sec:ca_in_practice}). This grounds the abstract property of changeability in practical mechanisms and limitations.
\end{enumerate}

The remainder of this paper is structured as follows: 
Section~\ref{sec:methodology} outlines our systematic review methodology. 
Section~\ref{sec:results} presents the results, including our categorization, definitions, layer model and related concepts. 
Section~\ref{sec:ca_in_practice} applies these insights to practical examples, while
Section~\ref{sec:discussion} discusses their broader implications. Finally,
Section~\ref{sec:threats_to_validity} explores possible threats to the validity of
our studies and Section~\ref{sec:conclusion} concludes.

\section{Research Methodology}
\label{sec:methodology}
In this section, we will detail our research protocol, the search and selection process, data extraction, analysis, and synthesis. 
Fundamentally, we follow the PRISMA 2020 guidelines~\cite{Prisma2020} for systematic literature reviews in selecting the sources for this systematization of knowledge.

\subsection{Research Protocol}
\label{sec:research_protocol}
The purpose of the research protocol is to ensure quality by specifying the research questions, defining inclusion and exclusion criteria for the final paper set selection, and applying the quality assurance plan throughout the process.
Our study addresses the following four research questions: 
\begin{description}
\item[RQ1:] \emph{How does the current literature define cryptographic agility and which relevant characteristics of cryptographic agility can we extract?}
\begin{quote}
    RQ1 examines the various definitions of cryptographic agility in academic and gray literature, and addresses the lack of consensus on its meaning. 
    It identifies shared aspects of the definitions and clusters them into coherent categories.
\end{quote}
\item[RQ2:] \emph{What are the defining characteristics of cryptographic agility and to which definition can we synthesize them based on the literature?}
\begin{quote}
    RQ2 synthesizes the results from RQ1 into a literature-based definition of cryptographic agility that unifies the most significant characteristics extracted from the literature.
\end{quote}
\item[RQ3:] \emph{What is a sharpened, context-independent definition of cryptographic agility that goes beyond existing literature?}
\begin{quote}
    RQ3 refines and consolidates the concept of cryptographic agility beyond the definition synthesized from the literature.
    Building on the results of RQ2, we revisit the key characteristics of cryptographic agility and derive a sharpened and context-independent definition of cryptographic agility.
\end{quote}
\item[RQ4:] \emph{What are the contextual domains of cryptographic agility and how
	can its relationship to related concepts be clarified?}
\begin{quote}
    RQ4 investigates how cryptographic agility manifests in different contexts and
		synthesizes them into the overarching domains of conceptual, software, hardware,
		and organization. We also clarify the relationship of cryptographic agility to the neighboring concepts of
		cryptographic versatility and cryptographic interoperability, in order to sharpen conceptual boundaries and avoid terminological ambiguity.
\end{quote}
\end{description}

Our selection criteria (Table~\ref{tab:selection_criteria}) function as a structured quality control measure to systematically assess articles based on predefined standards. Articles are incorporated into the final selection only if they satisfy all inclusion criteria and are free from any exclusionary conditions.
\begin{table}[tb]
    \small
        \caption{Selection Criteria.}
        \label{tab:selection_criteria}
        \raggedright
				\begin{tabularx}{\linewidth}{X}
          \toprule
          \textbf{Inclusion Criteria} %\tabularnewline
          %\midrule
          \begin{enumerate}
            \item Primary and secondary studies that explicitly provide their own definition of cryptographic agility or a composite definition based on already existing ones.
            \item Studies that are classified as academic literature or specific gray literature, including work in progress, technical and research reports, journal papers, preprints, white papers, conference or workshop contributions, government documents, blog posts, websites, RSS feeds, PhD and master theses.
          \end{enumerate} \tabularnewline~\tabularnewline
          \toprule
          \textbf{Exclusion Criteria} %\tabularnewline
          %\midrule
          \begin{enumerate}
            \item Duplicates of already included studies.
            \item Older versions of an already included study.
            \item Gray literature that are videos, podcasts, and webinars.
            \item Studies that directly quote another paper's definition and therefore do not provide their own definition of cryptographic agility.
            \item Studies that are not available, and hence not analyzable (e.g., the link to a web page is broken or the full text of an article is not accessible).
            \item Studies written in any language other than English.
          \end{enumerate} \tabularnewline
          \bottomrule
    \end{tabularx}
\end{table}

In the quality assurance process, the first author took the lead in defining each
stage, while the other authors reviewed and discussed it until a consensus was reached.
Specifically, Table~\ref{tab:quality_assurance_plan_table} shows the four stages of our quality assurance plan.
It ensured that all researchers were involved in each stage, which helped build a common understanding and prevent potential individual bias.
\definecolor{verylightgray}{gray}{0.96}
\begin{table}[tb]
  \centering
  \caption{Quality Assurance Plan.\label{tab:quality_assurance_plan_table}}
  \begin{tabularx}{\columnwidth}{@{}l!{\color{lightgray}\vrule}X@{}}
    \toprule
    \textbf{Stage} & \textbf{Lead Author Role} \\
    \midrule
    \textbf{Research Questions}           & defines questions \\
    \rowcolor{verylightgray}
    \textbf{Search Strategy}              & selects search engines, terms, and steps \\
    \textbf{Inclusion/Exclusion Criteria} & selects main criteria \\
    \rowcolor{verylightgray}
    \textbf{Category Assignment}          & assigns definitions to categories \\
    \bottomrule
  \end{tabularx}
\end{table}

\subsection{Search and Selection}
\label{sec:methodology:search_and_selection}
We implemented a structured search process for articles in predefined sources.
The final paper set consists of the most relevant literature, which is the result
of applying our selection criteria to the individual articles.
For our search, we chose IEEE Xplore, ACM Digital Library, Google Scholar and the Google search engine as resources.
All four are established and efficient search engines for finding peer-reviewed and gray
literature. The utilization of both peer-reviewed and non-peer-reviewed literature
enhances the diversity of the final paper set and ensures that definitions from
academia and practice are included. This practice is in line with Paez~\cite{Paez17GrayLiterature} who states that
the inclusion of gray literature can lead to less publication bias, enhanced comprehensiveness and a balanced provision of evidence.
Because non-peer-reviewed sources vary in rigor, we did not weight sources by perceived quality but let the selection criteria act as a uniform quality gate.
Inclusion Criterion~1 admits only sources with substantive definitional content and excludes passing mentions, Exclusion Criterion~4 removes sources that merely quote another definition, and Exclusion Criterion~3 excludes the weakest gray formats such as videos, podcasts, and webinars.
We consider all sources that were not peer-reviewed as gray literature.
For Google Scholar we used the query \texttt{(cryptograph* agil*) OR (agil* cryptograph*)}; for Google Search, IEEE Xplore, and the ACM Digital Library we used \texttt{crypto* agil*} (see replication package~\cite{replication_package}).

After our search, we found 84 sources with potential definitions of cryptographic agility based on the title, abstract, and content of the source.
This initial set contains a broad range of sources, including peer-reviewed articles, white papers, books, and standards.
We then applied our selection criteria (see~Table~\ref{tab:selection_criteria}) to each of the 84 articles, resulting in a final set of 48 articles, nine of which contain more than one definition.
Sources were included if they satisfied all inclusion criteria and none of the exclusion criteria.
Inclusion Criterion 1 is specific to our research questions and requires that a source explicitly provide its own definition of cryptographic agility or a composite definition based on existing ones.
Related terms such as cryptographic transition or cryptographic migration are cryptographic in nature but do not necessarily concern agility, whereas algorithm agility need not be cryptographic at all.
Sources using such terms were therefore included only if they satisfied this criterion, and were not treated as separate search targets.
The remaining criteria ensure methodological rigor by excluding duplicates, outdated versions, inaccessible sources, and non-English studies.
Disagreements during the application of these criteria were resolved through discussion among all authors until a shared interpretation was reached.

\subsection{Data Extraction, Analysis, and Synthesis}
\label{sec:methodology:data_extraction_analysis_synthesis}
We now describe our procedure for extracting, analyzing, and synthesizing the data. Our entire process can be traced and reproduced in detail with the help of our replication package, which contains all tables and mappings \cite{replication_package}.

To answer RQ1, we iteratively developed six distinct categories to classify the relevant aspects mentioned in the selected definitions and then assigned the extracted aspects of each definition to these categories.
Where cases were unclear, the authors discussed them until agreement was reached.
To support classification reliability, we refined the categories iteratively until every aspect of every reviewed definition could be mapped without residual cases.
The complete source-by-source mapping is published in the replication package~\cite{replication_package} for independent inspection.
This iterative mapping unifies heterogeneous terminology and consolidates overlapping perspectives.
We further enforce a consistent terminology via a mapping table (e.g.,~\emph{crypto $\rightarrow$ cryptography}, \emph{updating $\rightarrow$ update}, and normalized related terms such as \emph{information technology systems/information systems/deployed systems $\rightarrow$ systems}).
This resulted in a comprehensive and consistent assignment for each definition.

To answer RQ2, we systematically evaluated the categorized data, yielding six categories with between 14 and 31 unique values.
We then clustered these values into subcategories to reach a higher level of abstraction (see~Table~\ref{tab:results:clusters}).
Based on this synthesis, we formulated a literature-based definition of cryptographic agility (cf.~Definition~\ref{box:literature-based_def_of_ca}).

To answer RQ3, we revisited the key characteristics derived in RQ2 and refined them to exclude context-dependent elements, resulting in a sharpened, context-independent definition (cf.~Definition~\ref{box:context-independent_def_of_ca}).

To answer RQ4, we analyzed how cryptographic agility manifests across contexts and
synthesized them into four overarching domains—conceptual, software, hardware, and organization.
We also clarify the relationship of cryptographic agility to related concepts, namely cryptographic versatility (cf.~Definition~\ref{box:def_of_versatility}) and cryptographic interoperability (cf.~Definition~\ref{box:def_of_interoperability}).

\section{Results}
\label{sec:results}
This section contains an analysis and synthesis of our findings.
First, we outline descriptive statistics of the selected articles in
Section~\ref{sec:results:descriptive_statistics}. Then, we answer our
four research questions about the literature's variety in definitions in
Section~\ref{sec:results:variety_of_definitions}, extract a literature-based
definition in
Section~\ref{sec:results:context-independent_def_of_ca}, provide a clearer
context-independent definition in
Section~\ref{sec:results:context-independent_def_ca}, and explore the contextual
domains of cryptographic agility and its relation to other relevant concepts in
Section~\ref{sec:results:rq3_relationships_of_cryptographic_agility}.

\subsection{Descriptive Statistics}
\label{sec:results:descriptive_statistics}
This section presents the descriptive statistics of the selected sources, i.e.,~their
publication year, the types of literature, the authors' affiliation, and the
validators. A breakdown of this data can be found in Table~\ref{t:descriptive_stats}.
\definecolor{verylightgray}{gray}{0.96}
\begin{table}
\small
  \renewcommand{\cellalign}{tl}
  \centering
  \caption{Classification of all papers according to our descriptive statistics
    criteria.\label{t:descriptive_stats}}
  \begin{tabularx}{\columnwidth}{c!{\color{lightgray}\vrule}c!{\color{lightgray}\vrule}cc!{\color{lightgray}\vrule}ccc!{\color{lightgray}\vrule}ccc}
    \toprule
    \textbf{Ref} & \textbf{Year} &
    \multicolumn{2}{c!{\color{lightgray}\vrule}}{\textbf{Literature}} &
    \multicolumn{3}{c!{\color{lightgray}\vrule}}{\textbf{Author Affil.}} &
    \multicolumn{3}{c}{\textbf{Validator}} \\
    \midrule
      & &
      \rotatebox[origin=l]{90}{Peer-Reviewed} &
      \rotatebox[origin=l]{90}{Gray} &
      \rotatebox[origin=l]{90}{Academia} &
      \rotatebox[origin=l]{90}{Industry} &
      \rotatebox[origin=l]{90}{\parbox{2cm}{Stand. Body}} &
      \rotatebox[origin=l]{90}{3rd Party} &
      \rotatebox[origin=l]{90}{Committee} &
      \rotatebox[origin=l]{90}{Author(s)} \\
  \midrule
    \citer{Grote19ReviewOfPQC} & 2019 & \textbullet &  & \textbullet &  &  & \textbullet &  &  \\
    \rowcolor{verylightgray}
    \citer{Richter21AgileAndVersatileQuantum} & 2021 & \textbullet &  & \textbullet &  &  & \textbullet &  &  \\
    
    \citer{Badertscher22ComposableApproach} & 2022 & \textbullet &  & \textbullet & \textbullet &  & \textbullet &  &  \\
    \rowcolor{verylightgray}
    \citer{Badari21OverviewBitcoin} & 2021 &  & \textbullet & \textbullet &   \textbullet &  &  &  &  \textbullet \\
    
    \citer{Ma21CARAF} & 2021 & \textbullet &  &  & \textbullet &  & \textbullet &  &  \\
    \rowcolor{verylightgray}
    \citer{Moustafa18CAMustHave} & 2018 &  & \textbullet &  & \textbullet &  &  &  & \textbullet \\
    
    \citer{rfc6421} & 2011 &  & \textbullet &  & \textbullet & \textbullet &  & \textbullet &  \\
    \rowcolor{verylightgray}
    \citer{Nationalacademies17CAandInteroperability} & 2017 &  & \textbullet & \textbullet & \textbullet & \textbullet &  & \textbullet &  \\
    
    \citer{MacaulayCAinPractice} & 2019 &  & \textbullet &  & \textbullet &  &  &  & \textbullet \\
    \rowcolor{verylightgray}
    \citer{ETSI20CYBERMigrationStrategies} & 2020 &  & \textbullet &  &  & \textbullet &  & \textbullet &  \\
    
    \citer{Sikeridis23Enterprise-levelCA} & 2023 &  & \textbullet &  & \textbullet &  &  &  & \textbullet \\
    \rowcolor{verylightgray}
    \citer{Barker21GettingReadyForPQC} & 2021 &  & \textbullet &  &  & \textbullet &  & \textbullet &  \\
    
    \citer{rfc7696GuideslinesCA} & 2015 &  & \textbullet &  & \textbullet & \textbullet &  & \textbullet &  \\
    \rowcolor{verylightgray}
    \citer{Ott19ResearchChallenges} & 2019 & & \textbullet & \textbullet & \textbullet &  &  & \textbullet &  \\
    
    \citer{Fan21ImpactOfPQCHybridCerts} & 2021 & \textbullet &  & \textbullet & \textbullet &  & \textbullet &  &  \\
    \rowcolor{verylightgray}
    \citer{Vasic16LightweightSolution} & 2016 & \textbullet &  & \textbullet &  &  & \textbullet &  &  \\
    
    \citer{Zhang23MakingSWQuantumsafe} & 2023 & \textbullet & & \textbullet & \textbullet &  & \textbullet &  &  \\
    \rowcolor{verylightgray}
    \citer{ISARA20ManagingCryptoRisk} & 2020 &  & \textbullet &  & \textbullet &  &  &  & \textbullet \\
    
    \citer{Wiesmaier21PQCMigration} & 2021 &  & \textbullet & \textbullet &   \textbullet &  &  &  & \textbullet \\
    \rowcolor{verylightgray}
    \citer{Paul19ImportanceOfCA} & 2019 & \textbullet &  & \textbullet &  &  & \textbullet &  &  \\
    
    \citer{Alnahawi23StateOfCA} & 2023 &  & \textbullet & \textbullet &  &  &  &  & \textbullet \\
    \rowcolor{verylightgray}
    \citer{Paul21TransitionToPQCIoT} & 2021 &  & \textbullet & \textbullet &  &  & \textbullet &  &  \\
    
    \citer{Würth21PQC} & 2021 &  & \textbullet & \textbullet & \textbullet &  &  &  & \textbullet \\
    \rowcolor{verylightgray}
    \citer{Cunningham21SystemAcquisition} & 2021 &  & \textbullet & \textbullet &  &  &  &  & \textbullet \\
    
    \citer{digicert20PQCMaturityModel} & 2020 &  & \textbullet &  & \textbullet &  &  &  & \textbullet \\
    \rowcolor{verylightgray}
    \citer{Heftrig22PosterDNSSEC} & 2022 & \textbullet &  & \textbullet &   \textbullet &  & \textbullet &  &  \\
    
    \citer{Vasic12SecurityAgility} & 2012 & \textbullet &  & \textbullet &  &  & \textbullet &  &  \\
    \rowcolor{verylightgray}
    \citer{Sullivan09CA} & 2009 &  & \textbullet &  & \textbullet &  &  &  & \textbullet \\
    
    \citer{LaMacchia21LongRoadToPQC} & 2022 & \textbullet &  &  & \textbullet &  & \textbullet &  &  \\
    \rowcolor{verylightgray}
    \citer{Mashatan21ComplexPath} & 2021 & \textbullet &  & \textbullet &   \textbullet &  & \textbullet &  &  \\
    
    \citer{Mehrez18CAProperties} & 2018 & \textbullet &  &  & \textbullet &  & \textbullet &  &  \\
    \rowcolor{verylightgray}
    \citer{handbook} & 2023 &  & \textbullet & \textbullet & \textbullet & \textbullet &  & \textbullet &  \\
    
    \citer{Hohm23MaturityModelCA} & 2023 & \textbullet &  & \textbullet &  &  & \textbullet &  &  \\
    \rowcolor{verylightgray}
    \citer{Yunakovsky21TowardsSecurityRecommendations} & 2021 & \textbullet &  &  & \textbullet &  & \textbullet &  &  \\
    
    \citer{Heid23TracingCA} & 2023 & \textbullet &  &  & \textbullet &  & \textbullet &  &  \\
    \rowcolor{verylightgray}
    \citer{Henry18CA} & 2018 &  & \textbullet &  & \textbullet &  &  &  & \textbullet \\
    
    \citer{Ott23ResearchOnCryptographicTransition} & 2023 & \textbullet &  & \textbullet & \textbullet &  & \textbullet &  &  \\
    \rowcolor{verylightgray}
    \citer{marchesi2025CAinLightOfQuantumResistenec} & 2025 & \textbullet & & \textbullet & \textbullet & & \textbullet & & \\

    \citer{cho2024SWDefinedCryptographyDesignFeature} & 2024 & & \textbullet & & \textbullet & & & & \textbullet \\
    \rowcolor{verylightgray}
    \citer{Silonosov2024CAManifestoinE2EE} & 2024 & \textbullet & & \textbullet & & & \textbullet & & \\

    \citer{Alqabandi2024CAinMobileBanking} & 2024 & & \textbullet & & \textbullet & & & & \textbullet \\
    \rowcolor{verylightgray}
    \citer{ras2025CAinHWPhoenix} & 2025 & & \textbullet & \textbullet & & \textbullet & & & \textbullet \\

    \citer{frauenschlaeger2024CAinOperationalTech} & 2024 & \textbullet & & \textbullet & & & \textbullet & & \\
    \rowcolor{verylightgray}
    \citer{fries2024CAwithAttributeCerts} & 2024 & \textbullet & & & \textbullet & & \textbullet & & \\

    \citer{galambosHWsupportedCAinQKD} & 2025 & \textbullet & & \textbullet & & & \textbullet & & \\
    \rowcolor{verylightgray}
    \citer{sanon2025QuantumReadyMobileCommunications} & 2025 & \textbullet & & \textbullet & & & \textbullet & & \\

    \citer{kourtis2025AdaptivePQCforBlockchain} & 2025 & \textbullet & & \textbullet & & & \textbullet & & \\
    \rowcolor{verylightgray}
    \citer{barker2025NISTConsiderationsforCA} & 2025 & & \textbullet & \textbullet & \textbullet & \textbullet & & \textbullet & \\

    \midrule
    \midrule
    $\sum$ &  & 24 & 24 & 29 & 31 & 8 & 25 & 8 & 15 \\
  \bottomrule
  \end{tabularx}
\end{table}

\subsubsection{Publication Year}
\label{sec:results:year}
As Table~\ref{t:descriptive_stats} shows, the number of publications has almost
doubled since
2017 compared to previous years. There was a particular peak in 2021 with eleven
publications, followed by 2025 with six. This trend illustrates a significant
increase in interest in discussions of the term cryptographic agility.

\subsubsection{Type of Literature}
\label{sec:results:literature}
The type of literature is our second criterion and provides an initial overview
of the sources' provenance. We distinguish between peer-reviewed
and gray literature. The former in this context means that the paper was
part of a scientific publication process, including a peer review. We consider
all other kinds of sources, including preprints and RFCs, as gray literature.
For our study, the definitions themselves are more important than
whether they have been reviewed by a scientific audience, so we
also include gray literature. We draw upon data from 24 sources of gray literature and 24
peer-reviewed publications.

\subsubsection{Author Affiliation}
\label{sec:results:author_affiliation}
The affiliation can potentially influence the credibility and the reach of the sources. We
identified the following categories: Academia, Industry, and Standardization
Bodies. The authors of the selected sources can belong to one or more of these
three categories. Authors associated with industry form
the category with the most selected sources (31), followed by authors belonging to
academia (29). Authors affiliated with standardization bodies are the least
represented (8).

\subsubsection{Validator}
\label{sec:results:validator}
Our final criterion is ``validators'', which indicates how the authors of the
selected sources chose to validate their results. We found three categories:
\begin{description}
	\item[Third Party:] A team of independent experts who are not affiliated with the
authors of the original source. 
	\item[Committee:] A group of experts in the field who do not
necessarily need to be independent, i.e.,~the source's authors can be members of the
committee. 
	\item[Authors:] The authors of the original source validated their own findings without external input.
\end{description}
A source can only be assigned to one of the three validators. If the
validator of a source is not apparent, it is assigned to the Author category.
A total of 25 sources were assigned to third parties, eight to committee and 15 to authors.

\subsection{RQ1: Variety of Definitions}
\label{sec:results:variety_of_definitions}
By answering this research question, we aim to highlight the diversity of existing
definitions for the term cryptographic agility and structure the current state of
knowledge in order to provide a foundation for consensus.

Not all papers we considered for our review actually give their own, clear-cut definition. 
Instead, some just focus on specific subareas of cryptographic
agility~\citer{rfc6421,rfc7696GuideslinesCA,Vasic16LightweightSolution,Paul21TransitionToPQCIoT,Vasic12SecurityAgility},
like the agility of cryptographic
protocols~\citer{Vasic16LightweightSolution,Vasic12SecurityAgility}. Other papers
discuss relevant aspects of cryptographic agility without holistically defining it
themselves. They either just quickly describe their understanding of the notion of
cryptographic
agility~\citer{Richter21AgileAndVersatileQuantum,Ma21CARAF,Moustafa18CAMustHave,Barker21GettingReadyForPQC,Cunningham21SystemAcquisition}
or they rely on other
research~\citer{Badertscher22ComposableApproach,Badari21OverviewBitcoin,Zhang23MakingSWQuantumsafe,Wiesmaier21PQCMigration,Alnahawi23StateOfCA}.
This approach is taken by the largest group in our reviewed paper set. Sikeridis et
al.~\citer{Sikeridis23Enterprise-levelCA} even do both by first mentioning their
rough understanding and then elaborating on other sources' definitions, but without
synthesizing the different perspectives.

Apart from the previously mentioned papers, we found several attempts at forging a
holistic and context-independent definition of cryptographic agility. Those include:

``True and complete cryptographic agility is the ability to implement, update,
change, and remove cryptographic functions from systems and applications on demand,
without changing the systems or applications themselves. Cryptographic agility is
required across the spectrum of devices, applications, and systems we use today as
consumers and business, because cryptography itself is everywhere!\\ Cryptographic
agility should also include the notion of policy management [as] another partial form
of agility which allows rules to be applied around what type of crypto might be used
by an application or system, relative to what is available on the
system.''~\citer{MacaulayCAinPractice}

``In many ways, cryptographic agility represents the generalization of PQC migration
in that it considers not just the current challenge of migrating from our current
algorithms to PQC alternatives, but the long term need for ongoing migrations as new
attacks and better algorithms motivate the need for updates in our cryptographic
standards.''~\citer{Ott19ResearchChallenges}

``Cryptographic agility is the ability of a system to migrate easily from one
cryptographic algorithm to another, in a way that is flexible, scalable, and
dynamic.''~\citer{Mehrez18CAProperties}

The attempt given by Mehrez and Omri~\citer{Mehrez18CAProperties} is more formal than
most papers and resembles the common understanding that cryptographic agility has to
do with the ability to migrate and also includes appropriate quality attributes such
as ``flexible'' and ``dynamic''. Ott et al.~\citer{Ott19ResearchChallenges} provide a
detailed view on the topic of cryptographic agility. They themselves see their
description rather as a working definition for identifying relevant research
questions instead of being a holistic definition itself. We especially see their
focus on the post-quantum setting as too narrow for the more general notion of
cryptographic agility. Finally, Macaulay and Henderson~\citer{MacaulayCAinPractice}
forged a detailed and comprehensive, yet colloquial definition of cryptographic
agility. This description, again, lacks a clear, consistent explanation of the scope
and cryptographic entities to which it applies. For instance, for cryptographic
entities, it refers to applications and systems but once also to devices. These
inconsistencies make the definition less comprehensible to us. Also, we wonder
whether the given lists are meant to be exhaustive. For example, we would also see
infrastructures as an entity which can be cryptographically agile.

In summary, almost all the reviewed descriptions see cryptographic agility as related
to something that facilitates cryptographic migration, as stated in particular by Mehrez and
Omri~\citer{Mehrez18CAProperties}. However, the literature is in disagreement
regarding the relation of cryptographic agility to other aspects.
Those include whether cryptographic agility is also about protocols dynamically
negotiating their used algorithms (something we deem more related to interoperability
than agility, see
Section~\ref{sec:results:rq3_relationships_of_cryptographic_agility}), modular architecture of cryptography-related software, management
practices for cryptography, or whether it is a purely abstract concept. Also, the
level of abstraction is vastly different between the descriptions. In our three
examples we assessed one as being too generic~\citer{Mehrez18CAProperties}, one as
too narrow~\citer{Ott19ResearchChallenges}, and one as inconsistent in its level of
abstraction~\citer{MacaulayCAinPractice}. Hence, even though the content of many
definitions resembles our canonical one in large parts, those simplified definitions of
cryptographic agility do not reflect the entirety of the literature. Additionally,
we see these attempts as both falling short of holistically grasping the concept of
cryptographic agility and at the same time lacking clear-cut borders of what is and
what is not explicitly part of cryptographic agility.

To clarify this surprisingly complex situation, we applied a systematic
and iterative approach for classifying the variety of aspects of the
definitions. As a result, the following six distinct categories emerged. We give
a suitable question to classify the relevant aspects of a definition at the end
of each category:
\begin{description}
	\item[Context]{describes the setting in which cryptographic agility is
			applied in a given instance. In the literature, context ranges from
			very abstract terms, e.g.,~\emph{hardware} or \emph{software}, to
			somewhat more precise ones, like \emph{Internet of Things (IoT)}, \emph{cloud}, or
			\emph{programming languages} but may go down to a specific device or
			software component. Thus, the range of the context may be highly
			individual from abstract concepts like a strategy or processes down to
			very specific work on a concrete crypto instantiation in a software
			implementation. We believe context is highly relevant for finding specific
			measures to apply cryptographic agility in practice and the baseline
			for a closer examination.\\
	\emph{In which technological context is cryptographic agility mentioned?}}
	\item[Modes]{describe what cryptographic agility is seen as in the articles.
			For example, it varies from a \emph{cryptographic engineering} practice,
			to an \emph{approach} or an \emph{objective}. The mode can be seen
			as the foundation of how cryptographic agility is interpreted for
			the considered context.\\
	\emph{In which form or variety is cryptographic agility presented?}}
	\item[Desired Capabilities]{describe the extent to which capabilities are
			provided through cryptographic agility within a specific context, e.g.,~the capability of \emph{identifying} or \emph{modifying} the
			cryptographic entities of a system. \\
	\emph{What are the desired capabilities provided through cryptographic
	agility?}}
	\item[Quality Attributes]{further refine the desired capabilities, i.e.,~a more concrete
			description of properties and features.
			Possible examples are \emph{scalable}, \emph{easy}, or
			\emph{flexible}~\citer{Mehrez18CAProperties}. Inclusion of such
			information improves the quality of the definition by providing
			additional precision.\\
	\emph{Which quality attributes are provided through the desired
	capabilities?}}
	\item[Cryptographic Entities]{describe the different types and variants of the
			existing cryptography within the appropriate context. A cryptographic entity can include anything from keying material and encryption methods to parameter sets or even an entire crypto library. \\
		\emph{Which cryptographic entities are affected by cryptographic
		agility?}}
	\item[Drivers]{describe the real-world motivations or incentives behind
			cryptographic agility. For example, the necessary \emph{resilience
			against quantum attacks} or \emph{compromised classical algorithms}
			are typically mentioned by authors. Improved handling of new
			vulnerabilities, e.g.,~due to software bugs, is another example.\\
	\emph{What are motivations for realizing cryptographic agility?}}
\end{description}
After establishing six distinct categories, we continued our iterative mapping
procedure by assigning the definitions to the categories they consider.
Table~\ref{tab:category_coverage} shows the results of this process. Each of the
48 articles states some \emph{Desired Capabilities} of cryptographic agility,
and 46 also name the affected \emph{Cryptographic Entities}. In contrast, \emph{Drivers},
which motivate cryptographic agility, represent the least mentioned category with
only 31 articles. The other three categories \emph{Quality Attributes}
($n=39$\footnote{$n$ equals number of publications}), \emph{Mode} ($n=40$) and
\emph{Context} ($n=37$) are almost equally distributed. From that we deduce that
the \emph{Desired Capabilities} and \emph{Cryptographic Entities} are especially
important for a definition of cryptographic agility, while the \emph{Drivers} are
less of a defining factor. The other three categories seem to form the middle ground
in terms of significance for forming a definition.
\begin{table}
  \centering
  \caption{The articles' coverage of our categories.
  \label{tab:category_coverage}}
  \begin{tabularx}{\columnwidth}{YYYYYYY}
  \toprule
  \rotatebox{90}{\textbf{Ref}} & \rotatebox{90}{\textbf{Context}} &
  \rotatebox{90}{\textbf{Mode}} & \rotatebox{90}{\parbox{2cm}{\textbf{Desired\\
  Capabilities}}} &
  \rotatebox{90}{\parbox{2cm}{\textbf{Quality\\Attributes}}} &
  \rotatebox{90}{\parbox{2cm}{\textbf{Cryptographic Entities}}} &
    \rotatebox{90}{\textbf{Drivers}}
      \\ \midrule

  \citer{Grote19ReviewOfPQC} & \textbullet & & \textbullet &
       & \textbullet & \textbullet \\
  \rowcolor{verylightgray}
  \citer{Richter21AgileAndVersatileQuantum} &  & \textbullet & \textbullet & \textbullet & \textbullet & \textbullet \\
  \citer{Badertscher22ComposableApproach} &  & \textbullet & \textbullet & &\textbullet & \\
  \rowcolor{verylightgray}
  \citer{Badari21OverviewBitcoin} & &  & \textbullet & \textbullet & \textbullet & \\
  \citer{Ma21CARAF} & \textbullet & \textbullet & \textbullet & \textbullet & \textbullet & \textbullet \\
  \rowcolor{verylightgray}
  \citer{Moustafa18CAMustHave} & \textbullet & \textbullet & \textbullet & \textbullet & \textbullet & \textbullet \\
  \citer{rfc6421} & \textbullet & \textbullet & \textbullet & \textbullet & \textbullet & \textbullet \\
  \rowcolor{verylightgray}
  \citer{Nationalacademies17CAandInteroperability} & \textbullet & \textbullet & \textbullet & \textbullet & \textbullet & \textbullet \\
  \citer{MacaulayCAinPractice} & \textbullet & \textbullet & \textbullet & \textbullet & \textbullet &  \\
  \rowcolor{verylightgray}
  \citer{ETSI20CYBERMigrationStrategies} &  & \textbullet & \textbullet & & \textbullet  &  \\
  \citer{Sikeridis23Enterprise-levelCA} & \textbullet &  & \textbullet & \textbullet & \textbullet & \\
  \rowcolor{verylightgray}
  \citer{Barker21GettingReadyForPQC} & \textbullet & \textbullet & \textbullet & \textbullet & \textbullet & \textbullet \\
  \citer{rfc7696GuideslinesCA} & \textbullet & \textbullet & \textbullet & \textbullet & \textbullet & \textbullet \\
  \rowcolor{verylightgray}
  \citer{Ott19ResearchChallenges} &  & \textbullet & \textbullet & \textbullet & \textbullet & \textbullet \\
  \citer{Fan21ImpactOfPQCHybridCerts} & \textbullet &  & \textbullet & \textbullet & \textbullet & \textbullet \\
  \rowcolor{verylightgray}
  \citer{Vasic16LightweightSolution} & \textbullet & \textbullet & \textbullet & & \textbullet  & \textbullet \\
  \citer{Zhang23MakingSWQuantumsafe} & \textbullet & \textbullet & \textbullet & \textbullet & \textbullet &  \\
  \rowcolor{verylightgray}
  \citer{ISARA20ManagingCryptoRisk} & \textbullet & \textbullet & \textbullet & \textbullet & \textbullet & \textbullet \\
  \citer{Wiesmaier21PQCMigration} & \textbullet & \textbullet & \textbullet & & \textbullet  & \textbullet \\
  \rowcolor{verylightgray}
  \citer{Paul19ImportanceOfCA} & \textbullet & \textbullet & \textbullet & \textbullet & \textbullet & \textbullet \\
  \citer{Alnahawi23StateOfCA} & \textbullet & \textbullet & \textbullet & \textbullet & \textbullet &  \\
  \rowcolor{verylightgray}
  \citer{Paul21TransitionToPQCIoT} & \textbullet & \textbullet & \textbullet & \textbullet & \textbullet & \textbullet \\
  \citer{Würth21PQC} & \textbullet & \textbullet & \textbullet & \textbullet & \textbullet & \textbullet \\
  \rowcolor{verylightgray}
  \citer{Cunningham21SystemAcquisition} & \textbullet &  & \textbullet & \textbullet & \textbullet &  \\
  \citer{digicert20PQCMaturityModel} & \textbullet & \textbullet & \textbullet & \textbullet & \textbullet & \textbullet \\
  \rowcolor{verylightgray}
  \citer{Heftrig22PosterDNSSEC} & \textbullet & \textbullet & \textbullet & \textbullet & \textbullet & \textbullet \\
  \citer{Vasic12SecurityAgility} &  & \textbullet & \textbullet & & \textbullet  & \textbullet \\
  \rowcolor{verylightgray}
  \citer{Sullivan09CA} &  &  & \textbullet & \textbullet & \textbullet & \textbullet \\
  \citer{LaMacchia21LongRoadToPQC} & \textbullet & \textbullet & \textbullet & & \textbullet  & \textbullet \\
  \rowcolor{verylightgray}
  \citer{Mashatan21ComplexPath} & \textbullet & \textbullet & \textbullet & \textbullet & \textbullet &  \\
  \citer{Mehrez18CAProperties} & \textbullet & \textbullet & \textbullet & \textbullet & \textbullet &  \\
  \rowcolor{verylightgray}
  \citer{handbook} & \textbullet & \textbullet & \textbullet & \textbullet & \textbullet &  \\
  \citer{Hohm23MaturityModelCA} &  &  & \textbullet & \textbullet & \textbullet & \textbullet \\
  \rowcolor{verylightgray}
  \citer{Yunakovsky21TowardsSecurityRecommendations} & \textbullet & \textbullet & \textbullet & \textbullet & \textbullet & \textbullet \\
  \citer{Heid23TracingCA} & \textbullet & \textbullet & \textbullet & \textbullet & \textbullet & \textbullet \\
  \rowcolor{verylightgray}
  \citer{Henry18CA} & \textbullet & \textbullet & \textbullet & \textbullet & \textbullet & \textbullet \\
  \citer{Ott23ResearchOnCryptographicTransition} & \textbullet & \textbullet & \textbullet & & \textbullet  &  \\

  \rowcolor{verylightgray}
  \citer{marchesi2025CAinLightOfQuantumResistenec} & & \textbullet & \textbullet & \textbullet & \textbullet & \textbullet \\
  \citer{cho2024SWDefinedCryptographyDesignFeature} & \textbullet & \textbullet & \textbullet & \textbullet & \textbullet & \\

  \rowcolor{verylightgray}
  \citer{Silonosov2024CAManifestoinE2EE} & & \textbullet & \textbullet & \textbullet & & \textbullet \\
  \citer{Alqabandi2024CAinMobileBanking} & \textbullet & \textbullet & \textbullet & & & \textbullet \\

  \rowcolor{verylightgray}
  \citer{ras2025CAinHWPhoenix} & \textbullet & \textbullet & \textbullet & \textbullet & \textbullet & \textbullet \\
  \citer{frauenschlaeger2024CAinOperationalTech} & \textbullet & \textbullet & \textbullet & \textbullet & \textbullet & \\

  \rowcolor{verylightgray}
  \citer{fries2024CAwithAttributeCerts} & \textbullet & \textbullet & \textbullet & \textbullet & \textbullet & \textbullet \\
  \citer{galambosHWsupportedCAinQKD} & & & \textbullet & \textbullet & \textbullet & \\

  \rowcolor{verylightgray}
  \citer{sanon2025QuantumReadyMobileCommunications} & \textbullet & \textbullet & \textbullet & \textbullet & \textbullet & \textbullet \\
  \citer{kourtis2025AdaptivePQCforBlockchain} & \textbullet & \textbullet & \textbullet & \textbullet & \textbullet & \\

  \rowcolor{verylightgray}
  \citer{barker2025NISTConsiderationsforCA} & \textbullet & \textbullet & \textbullet & \textbullet & \textbullet & \\

  \midrule
  \midrule
  $\sum$ & 37 & 40 & 48 & 39 & 46 & 31 \\
  \bottomrule
\end{tabularx}
\end{table}

\\~\\
\begin{tcolorbox}[colback=gray!10,
	colframe=gray!50,
	title=\textbf{The Six Characteristics of Cryptographic Agility},
	before=\nolinebreak,
	coltitle=black]
  \label{box:six_categories_of_ca}
	Cryptographic Agility is a \emph{Mode} which provides \emph{Desired Capabilities} with specific \emph{Quality Attributes} concerning \emph{Cryptographic Entities} within a \emph{Context} motivated by \emph{Drivers}.
\end{tcolorbox}

\subsection{RQ2: Literature-Based Definition of Cryptographic Agility}
\label{sec:results:context-independent_def_of_ca}
To answer our second research question, we synthesize our findings from the
literature review based on the results from RQ1. We provide a comprehensive, literature-based definition for the term cryptographic agility which contains the essential characteristics that hold across all contexts.
This provides a common conceptual foundation, before reintroducing specific contexts
to capture how agility manifests in conceptual, software, hardware, or organizational levels.

Our first step towards formulating a definition is to determine which of the six
categories comprise defining attributes for cryptographic agility and which
categories only relate to it but are no defining factor.
This led us to exclude the categories \emph{Context} and \emph{Drivers}.
Context describes the setting in which cryptographic agility is applied and is therefore not a defining characteristic in itself.
Context is variable across settings, as what constitutes cryptographic agility in a software library differs from what it means in an organizational process.
Including context in the core definition would therefore produce a definition specific to one setting rather than a universally applicable one.
Drivers describe the motivations behind cryptographic agility, such as regulatory requirements, security threats, or algorithm obsolescence.
They explain why cryptographic agility is needed, but not what cryptographic agility is, and are therefore not a defining characteristic.
Excluding both allows a definition of cryptographic agility at its core that holds across contexts.
In contrast, the four categories \emph{Modes}, \emph{Desired Capabilities},
\emph{Quality Attributes}, and \emph{Cryptographic Entities} are the foundation of a proper definition.
\emph{Modes} describe what cryptographic agility actually is (e.g.,~a property) and
the \emph{Desired Capabilities} in combination with the \emph{Quality Attributes} what it accomplishes (e.g.,~efficient modification). Finally, the category \emph{Cryptographic Entities} describes the components affected by cryptographic agility.

After selecting the four relevant categories, we derived and unified the most
significant values for each category, shown in Table~\ref{tab:results:clusters}.
These values help answer the following three key questions that guide the establishment of a literature-based definition of cryptographic agility:
\begin{table}
	\caption{Clusters of Values for our definition of cryptographic agility.}
	\label{tab:results:clusters}
	\raggedright
	\begin{tabularx}{\columnwidth}{XX}
		\toprule
		\textbf{Category} & \textbf{Subcategories} \\
		\midrule
		Modes & Objective \\
		 & Property \\
		 & Approach (Theoretical, Practical) \\
		\midrule
		Desired Capabilities & Set up \\
		 & Identify \\
		 & Modify (Improve, Add, Remove, Exchange) \\
		\midrule
		Quality Attributes & Flexible \\
		 & Business-continuous \\
		 & Efficient (Time, Cost, Effort, Risk, Resource) \\
		\midrule
		Cryptographic Entities & Keying Material \\
		 & Encryption Methods (Parameters, Implementations, Standards) \\
		\bottomrule
	\end{tabularx}
\end{table}

\textbf{What are the unified modes of cryptographic agility?}
Categorizing the modes shows that cryptographic agility
is generally considered a (\emph{theoretical} or \emph{practical})
\emph{approach}, an \emph{objective}, or a \emph{property}. Frauenschlaeger et al. describe cryptographic agility as a ``dimension'' and ``aspect''~\citer{frauenschlaeger2024CAinOperationalTech}. 
We consider the underlying idea appropriate, yet find the terms too abstract and insufficiently precise for a unified and formal definition of cryptographic agility.

\textbf{What are the unified desired capabilities with their associated attributes
provided by cryptographic agility?}
As part of our synthesis, we identified three relevant subcategories for
\emph{Desired Capabilities}. These are the capabilities to \emph{set up},
\emph{identify}, or \emph{modify} cryptographic entities. We can divide \emph{modify}
further into \emph{improve}, \emph{add}, \emph{remove}, and \emph{exchange}.

The capabilities' quality attribute subcategories are \emph{flexible},
\emph{business-continuous}, and \emph{efficient} in terms of \emph{cost},
\emph{effort}, \emph{time}, \emph{risk}, and \emph{resources}. In our final results,
we deliberately exclude overly generic values, such as ``great'' or
``heterogeneous''.
We also considered the desired capabilities ``support''~\citer{Silonosov2024CAManifestoinE2EE} and ``coordination''~\citer{frauenschlaeger2024CAinOperationalTech} as misleading, since they describe auxiliary or interaction processes rather than distinct capabilities.
Hence, we also excluded these in order to create a formal definition.

\textbf{What are the unified cryptographic entities affected by cryptographic
agility?}
Based on our synthesis we derived the subcategories \emph{keying material} and
\emph{encryption method} for cryptographic entities. \emph{Encryption method} can
further be divided into \emph{parameters}, \emph{implementations}, and
\emph{standards}. Other aspects in the literature we appraise as too generic for
consideration were ``cryptography'', ``cryptographic core'', and ``cryptographic
features''.

With our subcategories as answers to the questions relevant for the definition at
hand, we can formulate the following literature-based definition of cryptographic
agility:
\\~\\
\begin{tcolorbox}[colback=gray!10,
	colframe=gray!50,
	title=\textbf{Literature-based Definition of Cryptographic Agility},
	before=\nolinebreak,
	coltitle=black]
  \label{box:literature-based_def_of_ca}
	\emph{Cryptographic Agility} is a theoretical or practical approach, objective, or property which provides capabilities for setting up, identifying, and
	modifying encryption methods and keying material in a flexible and efficient way
	while preserving business continuity.
\end{tcolorbox}

\subsection{RQ3: Context-Independent Definition of Cryptographic Agility}
\label{sec:results:context-independent_def_ca}
Building on the literature-based definition established in RQ2, we conducted an additional synthesis to further clarify the concept of cryptographic agility. 
In short, the literature-based definition (Definition~\ref{box:literature-based_def_of_ca}) combines the recurring characteristics found in prior work, whereas the context-independent definition (Definition~\ref{box:context-independent_def_of_ca}) reduces them to the single core property of changeability.
For this purpose, we revisited four key characteristics of the concept: its mode, desired capability, the role of quality attributes, and the notion of cryptographic entities.

First, we understand cryptographic agility as a \emph{property} inherent to a given context. 
In our synthesis of the literature, three possible modes emerged: property, approach, and objective.
We argue that \emph{property} is the most suitable. 
An \emph{objective} describes an external goal. This is misleading because
cryptographic agility is not a target to be achieved and then forgotten about but a
feature of the system under consideration.
In contrast, PQC migration could be considered as an objective since it represents a single, one-time event. 
An \emph{approach} describes a way of proceeding rather than a characteristic of the system itself. This is misleading because cryptographic agility does not depend on how change is pursued but on whether the system has the built-in property to support change.
A \emph{property}, in contrast, refers to an intrinsic feature of a system.
Describing cryptographic agility as a property makes it possible to specify and evaluate it clearly within a given context.

Second, we refine the understanding of the desired capabilities.
Based on our synthesis, cryptographic agility enables a context to \emph{set up} and \emph{modify} cryptography.
We unify these capabilities under the broader notion of \emph{changeability}.
Modification can be broken down into the fundamental operations \emph{add} and \emph{remove}, which define the system's ability to integrate new cryptographic entities and eliminate obsolete or insecure ones.
Replacement, often discussed in the literature, is a combination of these operations,
whereby one entity is removed and another added.
Configuration, expressed through \emph{enable} and \emph{disable}, represents the system's capacity to set up cryptography without altering its composition.
We highlight these four operations (\emph{add, remove, enable, disable}) because they
together capture the essence of what it means for a system to be changeable: They
describe the intrinsic mechanisms which allow altering a system's cryptographic entities.
Identification of cryptographic entities, while often mentioned in the literature, is
a requirement for change but not itself part of changeability.

Third, we deliberately exclude quality attributes from the definition.
Attributes such as flexibility, efficiency, or continuity describe how changeability is realized in a specific context.
They influence the performance or smoothness of cryptographic modifications but do not determine whether the property of cryptographic agility exists.
Including them confounds the definition with context-dependent aspects that vary between contexts.
By focusing on changeability itself, we define a property that exists independently of secondary characteristics, making the definition both context-independent and conceptually precise.
Excluding these attributes from the definition does not render them irrelevant.
Practical attributes such as cost, effort, risk, and operational continuity remain the criteria against which changeability must be assessed in a concrete context, as discussed in Sections~\ref{sec:discussion:complexity_of_ca} and~\ref{sec:discussion:assessment_of_cryptographic_agility}.

Fourth, we employ the term \emph{cryptographic entities} to refer to the affected cryptography and its components.
Compared to narrower notions such as keying material or encryption methods, the term cryptographic entities encompasses a wider spectrum.
Entities may include individual algorithms and keys, entire protocols, or cryptographic libraries, depending on the context.
This abstraction ensures that the property of cryptographic agility can be consistently recognized across different system layers and application scenarios, rather than being tied to specific implementations or artifacts.

Taking these refinements together, we propose the following final, context-independent definition of cryptographic agility as the answer to RQ3:
\\~\\
\begin{tcolorbox}[colback=gray!10,
  colframe=gray!50,
  title=\textbf{Context-Independent Definition of Cryptographic Agility},
  before=\nolinebreak,
  coltitle=black]
  \label{box:context-independent_def_of_ca}
  \emph{Cryptographic Agility} represents the changeability of cryptographic entities.
\end{tcolorbox}

\subsection{RQ4: Domains of Cryptographic Agility and its Related Concepts}
\label{sec:results:rq3_relationships_of_cryptographic_agility}
In this section, we move from the context-independent definition
(cf.~Definition~\ref{box:context-independent_def_of_ca}) of cryptographic agility
towards a more application-oriented perspective.
We first introduce a layer model to capture the various contexts of cryptographic agility and derive its overarching domains, namely the conceptual, software (cf. \citer{Nationalacademies17CAandInteroperability, MacaulayCAinPractice, Wiesmaier21PQCMigration, Ott23ResearchOnCryptographicTransition, kourtis2025AdaptivePQCforBlockchain}), hardware~(cf. \citer{Nationalacademies17CAandInteroperability, MacaulayCAinPractice, Wiesmaier21PQCMigration, Alnahawi23StateOfCA}), and organizational domain~(cf. \citer{Paul19ImportanceOfCA, Mashatan21ComplexPath, digicert20PQCMaturityModel}).
The model is derived from recurring contexts identified in the reviewed literature and should be understood as an open reference model.
Unlike assessment-oriented models such as CAMM~\citer{Hohm23MaturityModelCA}, it is descriptive: it organizes where cryptographic agility manifests rather than assigning maturity levels or scores.
Cryptographic agility spans all dimensions, cryptographic versatility is relevant in the hardware dimension, and cryptographic interoperability is relevant in the software and hardware dimensions, where cryptographic mechanisms are provided and used to exchange data securely across systems.
On this basis, we first introduce the layer model in detail before turning to the two related concepts of cryptographic versatility and interoperability.

\subsubsection*{Contextual Layers and Domains of Cryptographic Agility}
To clarify how cryptographic agility manifests across different environments, we
introduce a layer model that depicts the contexts of cryptography.
This model highlights that the property of changeability is not uniform, but appears
differently depending on the respective context.
For example, changes in the software domain such as libraries typically involve
substituting or updating algorithm implementations, whereas changes in the hardware
or organizational domain rather require procurement, governance, or supplier
management measures.
Based on our literature review, we identified 17 unique contexts, which are provided in our replication package.
By structuring these contexts, the model systematically illustrates where and how cryptographic agility arises and how dependencies between contexts affect its realization.

\begin{figure}[t]
  \centering
  \begin{tikzpicture}[
  scale=0.85, transform shape,
  primitive/.style      ={draw, ellipse, fill=gray!39, minimum width=2.0cm, minimum height=1.0cm, align=center},
  algorithm/.style      ={draw, ellipse, fill=gray!39, minimum width=3.0cm, minimum height=1.8cm},
  library/.style        ={draw, ellipse, fill=blue!60, minimum width=4.0cm, minimum height=2.6cm},
  application/.style    ={draw, ellipse, fill=blue!60, minimum width=5.0cm, minimum height=3.4cm},
  platform/.style       ={draw, ellipse, fill=blue!60, minimum width=6.0cm, minimum height=4.2cm},
  os/.style             ={draw, ellipse, fill=blue!60, minimum width=7.0cm, minimum height=5.0cm},
  device/.style         ={draw, ellipse, fill=green!47, minimum width=8.0cm, minimum height=5.8cm}, 
  infrastructure/.style ={draw, ellipse, fill=green!47, minimum width=9.0cm, minimum height=6.6cm},
  organization/.style   ={draw, ellipse, fill=red!48, minimum width=10.0cm, minimum height=7.4cm}
]

\node[organization, yshift=-3.2cm] (organization) {};
\node at ([yshift=-3.25cm]organization.center) {Organization};

\node[infrastructure, yshift=-2.8cm] (infrastructure) {};
\node at ([yshift=-2.86cm]infrastructure.center) {IT Infrastructure};

\node[device, yshift=-2.4cm] (device) {};
\node at ([yshift=-2.45cm]device.center) {Device};

\node[os, yshift=-2.0cm] (os) {};
\node at ([yshift=-2.05cm]os.center) {OS};

\node[platform, yshift=-1.6cm] (platform) {};
\node at ([yshift=-1.68cm]platform.center) {Platform};

\node[application, yshift=-1.2cm] (application) {};
\node at ([yshift=-1.28cm]application.center) {Application};

\node[library, yshift=-0.8cm] (library) {};
\node at ([yshift=-0.9cm]library.center) {Library};

\node[algorithm, yshift=-0.4cm] (algorithm) {};
\node at ([yshift=-0.45cm]algorithm.center) {Algorithm};

\node[primitive] (primitive) {Primitive};

\node[draw, rounded corners, fill=white, inner sep=4pt,
      anchor=north, yshift=-4mm] (legend)
  at (current bounding box.south) {
  \begin{tabular}{ll}
    \tikz{\node[primitive, rectangle, minimum width=6mm, minimum height=3mm]{};} &
		Conceptual Domain \\
    \tikz{\node[library, rectangle, minimum width=6mm, minimum height=3mm]{};} & Software Domain \\
    \tikz{\node[device,    rectangle, minimum width=6mm, minimum height=3mm]{};} & Hardware Domain \\
    \tikz{\node[organization, rectangle, minimum width=6mm, minimum height=3mm]{};} & Organizational Domain \\
  \end{tabular}
};

\begin{comment}
\begin{scope}[on background layer]
  \node[
    fill=none,
    draw=black!60,
    rounded corners=6pt,
    fit=(current bounding box),
    inner sep=10pt
  ] {};
\end{scope}
\end{comment}

\end{tikzpicture}
  \caption{Layer Model of Contexts in Which Cryptographic Agility Manifests Across Four Domains. Each Outer Layer Depends on and Is Constrained by the Layers Within It.}
  \label{fig:layer_model}
\end{figure}

\begin{table*}
  \centering
  \caption{Contexts of the Cryptographic Agility Layer Model with examples of how cryptographic agility is realized.}
  \label{tab:contexts_of_ca}
  \begin{tabular}{p{3cm}p{12.5cm}}
    \toprule
    \textbf{Layer} & \textbf{Examples of How Cryptographic Agility Is Realized} \\
    \midrule
    Primitives & Generic mathematical abstractions keep higher layers functional when the underlying structure changes (e.g.,~the same key exchange logic operates over finite-field, elliptic-curve, or lattice-based groups such as Module-LWE) \\
    \rowcolor{verylightgray}
    Algorithms & Parameterized configurations prevent code duplication and enable
		adaptable algorithm behavior (e.g.,~one AES implementation selecting the key
		length at runtime); shared validation layers enforce parameters and modes \\
    Libraries & Integration and substitution of cryptographic algorithm implementations (e.g.,~adding ML-KEM to OpenSSL) while retiring deprecated algorithms such as RC4 and MD5 \\
    \rowcolor{verylightgray}
    Applications & Configuration- or policy-based adaptation of cryptographic use (e.g.,~migrating messaging applications to post-quantum key exchange; disabling outdated TLS versions like 1.0/1.1) \\
    Platforms & Configurable replacement of platform components without disturbing dependent services (e.g.,~exchanging a web server such as NGINX for another within GitLab without influencing other components or services of the platform) \\
    \rowcolor{verylightgray}
    Operating System & Modular cryptographic frameworks and APIs allow algorithms to be updated or replaced (e.g.,~swapping legacy kernel crypto modules with PQC-capable ones; adding new primitives to the OS crypto API) \\ 
    Devices & Replacement or upgrade of hardware-based cryptographic components (e.g.,~exchanging TPM~1.2 for TPM~2.0; updating smartcards or hardware tokens to support post-quantum algorithms via firmware updates) \\
    \rowcolor{verylightgray}
    IT Infrastructure & Reconfiguration or replacement of infrastructure cryptographic services to adopt newer mechanisms (e.g.,~switching from RSA-based PKI to ECDSA- or PQC-based PKI; updating VPN and TLS termination systems) \\
    Organization & Adaptation of organizational policies and standards to evolving cryptographic requirements (e.g.,~adopting NIST post-quantum standards; updating procurement and compliance guidelines to enforce FIPS~140-3 conformity) \\
    \bottomrule
  \end{tabular}
\end{table*}

Figure~\ref{fig:layer_model} depicts our layer model.
Each layer represents a distinct context in which cryptographic entities are deployed
and also updated.
At the innermost level are cryptographic primitives---abstract building blocks such as block ciphers or hash functions---followed by algorithms that instantiate them and are typically implemented in software or hardware.
The model expands outwards to libraries, applications, platforms, operating systems, devices, IT infrastructure, and finally the organizational layer. 
This nested structure emphasizes dependencies: Outer layers build on inner ones and are thereby both enabled and constrained by them.
For example, applications depend on libraries or operating system frameworks that implement algorithms, while devices and infrastructure provide hardware-backed services under organizational governance.
Table~\ref{tab:contexts_of_ca} complements the layer model figure with representative
examples of cryptographic entities to be changed for each context.

\textbf{Domains as abstraction.}
We group the contexts into four overarching domains: the \emph{conceptual domain} (primitives, algorithms), the \emph{software domain}
(libraries, applications, platforms, operating systems), the \emph{hardware domain} (devices and infrastructure), and the \emph{organizational
domain}~(cf.~Figure~\ref{fig:layer_model}).
This grouping captures the most common and operationally distinct types of entities relevant for cryptographic agility, while enabling reasoning at a higher level of abstraction.
The distinction also explains the granularity of the model: 
The \emph{conceptual domain} represents the theoretical layer in which cryptographic
primitives and algorithms are defined, analyzed, and standardized independently of
any specific implementation. The contexts of this domain form the foundation for the
realizations on all
subsequent layers.
Software contexts are more fine-grained because cryptographic entities in software are highly modular, subject to frequent updates, and often nested (e.g.,~applications depending on other applications or libraries encapsulating multiple algorithms; cf.~Table~\ref{tab:contexts_of_ca}). 
Hardware contexts appear coarser, not because they are less complex but because
manufacturers of devices typically encapsulate several cryptographic components (e.g.,~accelerators, TPMs, enclaves; cf.~Table~\ref{tab:contexts_of_ca}) and infrastructures aggregate multiple devices.
At the organizational level, comparable subcontexts are less straightforward to
depict, since governance structures, processes, and product strategies are highly individual.
For this reason, Figure~\ref{fig:layer_model} does not visualize detailed organizational subcontexts and instead introduces representative examples in Table~\ref{tab:contexts_of_ca}.

\textbf{Representative, not exhaustive.}
The layer model is not intended as an exhaustive enumeration of all possible
contexts, but rather as a reference model highlighting the most representative ones consistently observed in literature and practice.
Additional contexts, such as cryptographic protocols, could be added in more detailed analyses. 
Protocols, for example, are typically implemented in software (as part of a library or application), may rely on hardware features (e.g.,~HSMs or TPMs), and are governed by organizational standards such as RFCs. 
This illustrates how changeability may simultaneously depend on factors of all
domains.
Hence, the model should be understood as open and extensible.

\textbf{Nested contexts.}
Even within a single domain, and sometimes even within a single layer, contexts may be nested. 
This is particularly visible in the software domain: platforms integrate applications, and applications can in turn contain or depend on other applications, nesting contexts within one layer. 
We illustrate this in Section~\ref{sec:ca_in_practice}, where GitLab CI/CD (a platform) internally integrates NGINX (an application), thereby forming a layered relationship within the software domain. 
Such nesting underlines that while the model provides a structured abstraction, concrete deployments often exhibit additional complexity.

\textbf{Implications for cryptographic agility.}
Each domain brings characteristic implications for the realization of cryptographic agility and thus provides a domain-specific perspective on its definition:

\emph{Conceptual domain.} Cryptographic agility in the conceptual domain denotes the property of changeability of abstract cryptographic entities, including primitives and algorithm specifications independent of their implementation.
It manifests primarily through the ability to introduce, modify, or deprecate cryptographic designs via research, standardization, or regulatory processes.
Although these changes are not operational by themselves, they determine the scope of agility achievable in all other domains.

\emph{Software domain.} Cryptographic agility in the software domain denotes the property of changeability of software-based cryptographic entities, including libraries, application bindings, and operating system providers.
It is shaped by modularity, interface stability, and update mechanisms.
Entities can often be substituted rapidly, but only if interfaces are well designed and backward compatibility is maintained.
A stable interface alone, however, is not enough: it can hide that a swapped scheme carries different functional guarantees (e.g.,~a different homomorphic structure or a stateful instead of a stateless signature scheme).
This exemplifies an inherent trade-off: the abstraction that enables swapping also conceals differences that callers may depend on.
OpenSSL's provider architecture keeps the API stable so algorithms can be swapped without changing application code, but ensuring that a swapped scheme still fulfills what the API promises remains the developer's responsibility, for example through interface contracts, conformance tests, and validation of the new provider~(cf.~Section~\ref{sec:ca_in_practice:openssl}).

\emph{Hardware domain.} Cryptographic agility in the hardware domain denotes the property of changeability of hardware-based cryptographic entities, including device components (e.g.,~accelerators, TPMs, HSMs), entire devices, and the infrastructures built upon them.
It is limited by physical constraints and lifecycle costs, since changes often require new procurement and depend on manufacturers’ support for updated cryptographic features.

\emph{Organizational domain.} Cryptographic agility in the organizational domain denotes the property of changeability of organizational entities, including cryptographic policies, standards, and suppliers that govern and shape the use of software and hardware components.
It is driven by governance, processes, and product lifecycles, which determine whether cryptographic changes can be integrated without disrupting operations.

\subsubsection*{Cryptographic Versatility}
\label{sec:results:rq4_cryptographic_versatility_def}
Having established the layer model and its domains, we now turn to related concepts that complement cryptographic agility. The first of these is \emph{cryptographic versatility}, originally introduced by Richter et al.~\citer{Richter21AgileAndVersatileQuantum} as the ability of a hardware system to perform multiple different cryptographic tasks. We consider this concept closely related to the hardware domain, where changeability is often limited by physical constraints.

In contrast to \emph{cryptographic agility}, which denotes the \emph{property} of changeability, versatility does not imply that new schemes can later be added or old ones removed.
Instead, it describes a static capability inherent to hardware. A versatile device
may, for instance, provide acceleration for AES, \mbox{SHA-2}, and ECC in parallel,
but it cannot adapt to support post-quantum cryptography unless this was anticipated
at design time. This ability may also include support for \emph{hybrid cryptographic
schemes}, where classical and post-quantum algorithms are combined to balance trust
in established primitives with protection against emerging quantum
threats~\cite{ietf-pquip-pqt-hybrid-terminology-04, MAJOT201517}. Versatility is
therefore not a sub-property of agility but a distinct ability that complements it in
contexts where hardware constraints inherently limit changeability.
\\~\\
\begin{tcolorbox}[colback=gray!10,
    colframe=gray!50,
    title=\textbf{Cryptographic Versatility},
    before=\nolinebreak,
    coltitle=black]
    \label{box:def_of_versatility}
    \emph{Cryptographic versatility} is the \emph{ability} of hardware entities to support multiple cryptographic schemes by design, including hybrid
    operations.
\end{tcolorbox}

\subsubsection*{Cryptographic Interoperability}
\label{sec:results:rq4_cryptographic_interoperability_def}
In addition to versatility, a second concept arises in the context of RQ4, namely \emph{cryptographic interoperability}.
Only a subset of the literature on cryptographic agility addresses it explicitly and
then without a uniform terminology: some authors refer to interoperability, others to compatibility, and some blur the line with agility.
This lack of precision has contributed to the frequent conflation of
both concepts. Historically, this conflation was common especially in the context of
protocols. There, the similarities between the common algorithms allowed protocols to
dynamically negotiate the exact cryptographic parameters for a connection. This made
them both interoperable and reduced the actual
agility to a software or library issue. However, the great variation in the characteristics
of the PQC algorithms means that they are usually not a
\mbox{drop-in} replacement for one another but that the context has to adapt to their specific features. Thus,
merging all the concepts together under the term of cryptographic agility is no longer
applicable. To properly define cryptographic interoperability for the current
situation, we first review how it is discussed in prior work and then contrast it
with agility.

Existing literature frames interoperability primarily as a requirement that accompanies cryptographic agility. 
McKay defines it broadly as ``the ability to communicate and exchange information between systems''~\citer{Nationalacademies17CAandInteroperability}. 
Mehrez and Omri~\citer{Mehrez18CAProperties} as well as the authors of
RFC6421~\citer{rfc6421} describe interoperability as a partial objective of agility,
since cryptographically agile solutions must also ensure communication across different implementations. 
Mehrez and Omri further emphasize that protocol designers carry responsibility for interoperability by specifying mandatory-to-implement algorithms. 
Other works associate interoperability with backward compatibility: Fan et al.\ highlight its role in integrating post-quantum cryptography without disrupting existing processes~\citer{Fan21ImpactOfPQCHybridCerts}, and RFC6421 underscores backward compatibility as a mandatory requirement in the RADIUS protocol~\citer{rfc6421}. 
Paul et al.\ note that newer protocol versions must continue to support secure communication across implementations~\citer{Paul21TransitionToPQCIoT}, while Moustafa et al.\ stress the importance of uniform standards as a prerequisite for interoperability~\citer{Moustafa18CAMustHave}. 
At the same time, several studies caution that excessive backward compatibility
undermines security by prolonging the use of deprecated
algorithms~\citer{MacaulayCAinPractice}\cite{dnssec_compat} and enabling downgrade
attacks~\cite{downgrade,TLS_version_downgrade}. These notions already show the high
cost in terms of complexity of interoperability, also compared with agility.
Furthermore, interoperability is still most frequently discussed in the context of protocols (e.g.,~\citer{Mehrez18CAProperties,Fan21ImpactOfPQCHybridCerts,rfc6421,Paul21TransitionToPQCIoT,Würth21PQC,Grote19ReviewOfPQC}).
Building on this foundation, we clarify the distinction between interoperability and agility and emphasize that the relationship between cryptographic agility and interoperability is unidirectional.

First, agility facilitates interoperability. 
Since agility encompasses the ability to add, enable, disable, and remove cryptographic entities, it enables systems to adapt their supported cryptographic mechanisms to the requirements of secure communication. 
For instance, if legacy support is necessary to maintain compatibility with outdated
systems, a cryptographically agile system can simply enable or add the relevant algorithms. 
If new secure algorithms must be adopted due to regulatory or security requirements, they can be integrated without disruption.
In both cases, these changes would increase the cryptographic interoperability of the system.
Conversely, when legacy support is no longer desired, agility also allows these
algorithms to be disabled or fully removed, increasing security by reducing interoperability.
This illustrates how interoperability depends on cryptographic agility. 
Furthermore, interoperability should be understood as tied to the set of active cryptographic mechanisms in a system. 
Even if a system could easily enable additional algorithms, these do not contribute to its current interoperability until it actually does so.

Second, cryptographic interoperability does not, by itself, determine cryptographic agility. 
The ability of a system to communicate securely with many peers through multiple
cryptographic mechanisms says nothing about the ease of changing its cryptography.
For example, supporting a wide range of algorithms does not automatically imply that they can be added, removed, or replaced efficiently.
Interoperability therefore only indicates the variety of secure communication possible at a given point in time, but neither enables nor constrains agility.
This distinction is particularly relevant considering cryptographic migration,
such as the upcoming PQC transition.
Migration typically aims at increasing security at the cost of interoperability by deprecating legacy schemes.
In such cases, agility is the decisive property, as it determines the efficiency of
disabling or removing obsolete algorithms from a system.

Cryptographic agility says nothing about which mechanisms are currently active.
Cryptographic interoperability, in turn, reflects exactly that: the set of mechanisms a system actively uses to communicate securely at a given point in time.
The two concepts are therefore distinct but related, and we refine the notion of interoperability as follows:
\\~\\
\begin{tcolorbox}[colback=gray!10,
	colframe=gray!50,
	title=\textbf{Cryptographic Interoperability},
	before=\nolinebreak,
	coltitle=black]
  \label{box:def_of_interoperability}
  \emph{Cryptographic Interoperability} is the ability of a system to
	communicate securely with others, using a variety of
	cryptographic mechanisms at the same time.
\end{tcolorbox}

\section{Cryptographic Agility in Practice}%
\label{sec:ca_in_practice}
In this section, we apply the concepts from Section~\ref{sec:results} to show how cryptographic agility manifests in specific contexts. 
The exemplars, i.e., OpenSSL, NGINX, and GitLab CI/CD, are widely deployed and maintained, which makes them relevant for real-world cryptographic agility considerations.
Throughout, we highlight each example's role within the layer model (cf.~Figure~\ref{fig:layer_model}), explain how cryptographic agility is realized and constrained, and assess its implications for interoperability (cf.~Section~\ref{sec:results:rq4_cryptographic_interoperability_def}).

\subsection{OpenSSL as a Cryptographically Agile Library}
\label{sec:ca_in_practice:openssl}
\begin{figure}[t]
  \centering
  \begin{tikzpicture}[
  scale=0.80, transform shape, >=Latex, font=\sffamily,%\normalsize,
  pill/.style    ={draw, ellipse, fill=white, minimum width=3.2cm, minimum height=9mm, align=center},
  bar/.style     ={draw, rounded corners=3pt,  fill=blue!55, minimum width=9.8cm, minimum height=6mm, align=center},
  provider/.style={draw, rounded corners=10pt, fill=gray!15, minimum width=4.0cm, minimum height=3.5cm},
  ptitle/.style  ={draw, rounded corners=6pt,  fill=blue!55, minimum width=3.0cm, minimum height=8mm, align=center},
  chip/.style    ={draw, rounded corners=6pt,  fill=gray!50, minimum width=1.35cm, minimum height=9mm, align=center},
  line/.style    ={->, thick, draw=gray!90}
]

% EVP-Leiste und Anwendung
\node[pill] (app) at (0,3.6) {Application};
\node[bar] (evp) at (0,2.2) {EVP-API};

% Provider-Container
\node[provider] (defp) at (-2.8,-0.5) {};
\node[provider] (legp) at ( 2.8,-0.5) {};

% Provider-Titel
\node[ptitle, anchor=north] (defp_ptitle) at (defp.north) {Default-Provider};
\node[ptitle, anchor=north] (legp_ptitle) at (legp.north) {Legacy-Provider};

% Kacheln im Default-Provider
\node[chip] at ([xshift=-1.0cm,yshift=-1.5cm]defp.north) {AES};
\node[chip] at ([xshift= 1.0cm,yshift=-1.5cm]defp.north) {RSA};
\node[chip] at ([xshift=-1.0cm,yshift=-2.75cm]defp.north) {EC};
\node[chip] at ([xshift= 1.0cm,yshift=-2.75cm]defp.north) {\dots};

% Kacheln im Legacy-Provider
\node[chip] at ([xshift=-1.0cm,yshift=-1.5cm]legp.north) {MD2};
\node[chip] at ([xshift= 1.0cm,yshift=-1.5cm]legp.north) {RC4};
\node[chip] at ([xshift=-1.0cm,yshift=-2.75cm]legp.north) {\dots};

% Ellipse-Punkte in der Mitte
\node at ($(defp.east)!0.5!(legp.west)$) {\Large \dots};

% Edges
%\draw[line] (app.south) -- (evp.north)
\draw[->, line width=0.65pt, draw=gray!70] (app.south) -- (evp.north);
\draw[line] ([xshift=-2.8cm]evp.south) -- (defp.north);
\draw[line] ([xshift= 2.8cm]evp.south) -- (legp.north);

\end{tikzpicture}
  \caption{OpenSSL Provider Architecture Centered on the EVP API. The EVP Abstraction Decouples Applications from Algorithm Implementations; Providers Can Be Added, Removed, or Replaced Without Recompiling Dependent Applications.}
  \label{fig:envelope-api}
\end{figure}
OpenSSL\footnote{\url{https://docs.openssl.org/master/man1/openssl/}} is a software library for cryptography that resides at the library layer of our layer model (cf.~Figure~\ref{fig:layer_model}).
It provides cryptographic primitives and algorithms to a broad spectrum of applications, ranging from embedded devices to large-scale web services. 
Because it sits at the interface between low-level algorithms and higher-level applications, its treatment of cryptographic agility strongly influences other layers.

\textbf{Realization of cryptographic agility.} 
OpenSSL realizes cryptographic agility primarily through its provider architecture, introduced with version~3.0.
The architecture is illustrated by Figure~\ref{fig:envelope-api}. 
Providers are modular containers for algorithm implementations that can be loaded dynamically at runtime. 
This design allows administrators or developers to add and remove cryptographic
algorithms without recompiling dependent applications, as long as those applications
rely on the unified Envelope (EVP) API\@.
The EVP API abstracts the details of individual algorithms by exposing a consistent interface for symmetric and asymmetric operations, digests, and key management. 
Through this abstraction, applications can switch algorithms by altering configuration or provider settings rather than modifying source code. 
Further agility is supported by OpenSSL's configuration system, which permits runtime selection of default providers, algorithm preferences, and security policies. 
Since version~3.5.0, the library has also included support for post-quantum algorithms, demonstrating its extensibility to entirely new cryptographic families. 
In summary, the provider architecture, the EVP API, and the runtime configuration
system are the key mechanisms that enable cryptographic agility in the context of
OpenSSL (cf.~OpenSSL~documentation\footnote{\url{https://docs.openssl.org}}%
~\footnote{\url{https://github.com/openssl/web/blob/master/docs/OpenSSL300Design.md}}%
~\footnote{\url{https://github.com/openssl/openssl/blob/master/README-PROVIDERS.md}}).

\textbf{Constraints of cryptographic agility.} 
Despite this modular design, OpenSSL faces structural limits. 
The EVP API must be used consistently; legacy applications that call lower-level APIs (e.g.,~directly invoking algorithm-specific functions) cannot benefit from the provider model and must be rewritten to gain cryptographic agility. 
Furthermore, OpenSSL lacks a unified mechanism for centralized cryptographic policy enforcement across multiple applications. 
Policies such as \emph{disallow RSA with key sizes below 2048 bits} must be configured separately for each consuming application, leading to potential misconfiguration. 
Dynamic loading of providers improves flexibility but does not guarantee the correctness or trustworthiness of a new implementation, which remains an operational and verification effort. 
Finally, integration of new cryptographic primitives often requires coordination with consuming applications to ensure correct parameter handling, especially in the case of emerging PQC algorithms with non-standardized interfaces.  
In summary, the main constraints on cryptographic agility in OpenSSL are legacy use of low-level APIs, lack of centralized policy management, and trustworthiness challenges with dynamically loaded providers.

\textbf{Interoperability.} 
OpenSSL directly impacts cryptographic interoperability by determining which algorithms and protocol versions are available to higher layers. 
For example, the cipher suites that a TLS implementation can negotiate depend
entirely on the algorithms exposed by OpenSSL\@.
However, because policy settings are decentralized, achieving consistent interoperability across a heterogeneous system landscape can be difficult.
In practice, interoperability is enabled by the library but still
mostly depends on the outer context, usually applications (cf.\
Figure~\ref{fig:layer_model}), to properly handle the
algorithms and maintain the policy settings.

\subsection{NGINX as a Cryptographically Agile Application}
\label{sec:ca_in_practice:nginx}
\begin{figure}[t]
  \centering
  \begin{tikzpicture}[
  scale=0.65, transform shape, >=Latex,
  node distance=10mm and 12mm, font=\sffamily,%\small,
  title/.style={font=\sffamily\large},
  app/.style   ={draw, rounded corners=3pt, fill=blue!55, minimum width=40mm, minimum height=9mm, align=center},
  box/.style   ={draw, rounded corners=3pt, fill=gray!50, minimum width=38mm, minimum height=8mm, align=center},
  circlebox/.style={draw, circle, fill=white, minimum size=15.5mm, align=center},
  line/.style  ={->, thick, draw=gray!70}
]

% Spaltenanker
\coordinate (L) at (-4.6,0);
\coordinate (R) at ( 4.6,0);

% Überschriften
\node[title, anchor=south] at ($(L)+(1.0,3.4)$) {Build-Time};
\node[title, anchor=south] at ($(R)+(-1.0,3.4)$) {Runtime};

% Trennlinie
\draw[gray!50] (0,3.6) -- (0,-2.2);

% ---------------- Build-Time (links)
\node[app, anchor=north] (bin) at ($(L)+(1.0,3.0)$) {NGINX Application};
\node[box, below=12mm of bin] (src) {NGINX Source Code};
\node[box, below=12mm of src] (ossl) {OpenSSL\\\small Crypto Library};

\draw[line] (ossl.north) -- node[pos=.5, fill=white, inner sep=1pt, below]
  {\small linked with} (src.south);
\draw[line] (src.north) -- node[pos=.5, fill=white, inner sep=1pt, below]
  {\small compiles} (bin.south);

% ---------------- Runtime (rechts)
\node[circlebox, anchor=north] (http) at ($(R)+(-1.0,3.0)$) {HTTPS};
\node[app, below=12mm of http] (rbin) {NGINX Application};
\node[box, below=12mm of rbin] (conf) {nginx.conf\\\small TLS Settings};

\draw[line] (http.south) -- (rbin.north);
\draw[line] (conf.north) -- node[pos=.5, fill=white, inner sep=1pt, below]
  {\small (hot-)reload} (rbin.south);

\end{tikzpicture}
  \caption{NGINX Cryptographic Agility at Build Time and Runtime. The Backend Is Fixed at Compile Time; TLS Settings Can Be Reconfigured and Reloaded Without Restarting.}
  \label{fig:nginx}
\end{figure}
NGINX\footnote{\url{https://nginx.org/}} is a high-performance, open-source web
server and reverse proxy that resides at the application layer of our layer model
(cf.~Figure~\ref{fig:layer_model}).
Although its core functionality is not limited to cryptography, it integrates cryptographic operations through external libraries such as OpenSSL and thereby inherits both strengths and weaknesses from the underlying library.

\textbf{Realization of cryptographic agility.} 
NGINX realizes cryptographic agility primarily through its configuration-driven design.
Administrators can enable, disable, or prioritize specific protocols, algorithms, and parameter sets directly in configuration files without modifying source code. 
At runtime, changes to TLS versions, ciphersuites, and key lengths can usually be applied by reloading the configuration, minimizing downtime. 
At compile-time, NGINX can be built against different cryptographic libraries or library versions, which allows operators to migrate to newer backends (e.g.,~OpenSSL with PQC support) without altering NGINX itself. 
This dual mechanism, namely runtime configuration for fine-grained control and compile-time linking for backend replacement, gives operators substantial flexibility in adapting cryptography to evolving requirements.  
In summary, cryptographic agility in NGINX is realized through its runtime configuration model and its ability to switch underlying cryptographic libraries at build time (cf.~NGINX~documentation\footnote{\url{https://nginx.org/en/docs/}}~\footnote{\url{https://nginx.org/en/docs/configure.html}}~\footnote{\url{https://github.com/nginx/nginx}}).

\textbf{Constraints of cryptographic agility.} 
NGINX's agility is constrained by its strong dependency on the capabilities of the linked library, most commonly OpenSSL. 
If the underlying library lacks support for a new primitive, NGINX cannot provide it. 
Furthermore, compile-time linking to a specific library version reduces flexibility, since migration to an alternative backend typically requires recompilation and sometimes code-level adjustments. 
Another constraint lies in the limited granularity of policy enforcement: while administrators can configure cipher priorities, enforcing organization-wide rules across many deployments requires external management tools.  
In summary, the main constraints on cryptographic agility in NGINX are its reliance on the underlying cryptographic library, the need for recompilation when switching backends, and limited native support for centralized policy enforcement.

\textbf{Interoperability.} 
NGINX plays a central role in establishing cryptographic interoperability for web services. 
Its configuration determines which TLS versions and ciphersuites can be negotiated with clients, directly impacting the range of peers it can communicate with. 
Enabling older algorithms improves backward compatibility but increases the risk of downgrade attacks, whereas disabling them raises security but may cut off legacy clients. 
Thus, NGINX illustrates how interoperability is shaped by cryptographic agility: changes in configuration or backend libraries translate into changes in the set of secure communication partners.

\subsection{GitLab CI/CD as a Cryptographically Agile DevSecOps Platform}%
\label{sec:ca_in_practice:gitlab}
\begin{figure}[t]
  \centering
  \begin{tikzpicture}[
  scale=0.65, transform shape, >=Latex,
  node distance     =10mm and 12mm, font=\sffamily,%\small,
  svc/.style        ={draw, rounded corners=2pt, fill=gray!50, minimum width=30mm,
	minimum height=8mm, align=center, anchor=center},
  circlebox/.style  ={draw, circle, fill=white, minimum size=15.5mm, align=center},
  line/.style       ={->, thick, draw=gray!70}
]

% Draw Other Boxes
\node[svc] (work)  {GitLab Workhorse};
\node[svc, left=of work] (pages) {GitLab Pages};
\node[svc, below=of work] (puma)  {Puma (GitLab Rails)};
\node[svc, right=of puma, minimum width=20mm] (gitaly) {Gitaly};

% Draw Nginx and GitLab Shell Boxes
\node[svc, fill=blue!55, above=of pages] (nginx) {NGINX};
\node[svc, fill=blue!55, above=24mm of gitaly]  (shell) {GitLab Shell};

% Draw Protocol Circles
\node[circlebox, above=of nginx] (http) {HTTP/\\HTTPS};
\node[circlebox, above=of shell] (ssh) {SSH};

% Draw Edges
\draw[line] (http) -- node[pos=.55, fill=white, inner sep=1pt, above]
  {\small TCP 80,443} (nginx);
\draw[line] (ssh)  -- node[pos=.55, fill=white, inner sep=1pt, above]
  {\small TCP 22} (shell);
\draw[line] (nginx) -- node[pos=.55, fill=white, inner sep=1pt, above]
  {\small TCP 8090} (pages);
\draw[line] (nginx.east) -- (work);
\draw[line] (shell.west) -- (work);
\draw[line] (work)  -- (puma);
\draw[line] (puma)  -- (gitaly);
\draw[line, <->] (work.east)  -- (gitaly);
\draw[line] (shell) -- (gitaly);

\end{tikzpicture}
  \caption{Cryptography-Relevant Components in GitLab CI/CD. NGINX Handles TLS Termination and GitLab Shell Manages SSH; Together They Form the Primary Cryptographic Endpoints of the Platform.}
  \label{fig:gitlab}
\end{figure}
GitLab~CI/CD\footnote{\url{https://about.gitlab.com}} is a widely used DevSecOps platform that we classify at the platform layer of our layer model (cf.~Figure~\ref{fig:layer_model}). 
As a platform, it integrates multiple services such as runners, registries, web servers, and repositories, which together provide continuous integration and deployment workflows. 
Its relevance for cryptographic agility lies in the way these services coordinate security-critical tasks like code signing, artifact integrity verification, and secure communication across the software delivery process.

\textbf{Realization of cryptographic agility.} 
Cryptographic agility in GitLab CI/CD is realized through its extensible pipeline model and the ability to integrate both internal and external cryptographic components. 
Internally, TLS termination and negotiation are handled via NGINX as a reverse proxy, while GitLab Shell manages SSH-based interactions. 
For artifact signing and integrity protection, GitLab integrates with tools such as GPG or Cosign. 
Externally, pipelines can invoke cryptographic libraries (e.g.,~OpenSSL), command-line utilities, or key management systems. 
This design makes it possible to add or replace cryptographic mechanisms without altering GitLab's core codebase. 
Security standards can be enforced directly in jobs, and key material can be rotated automatically via vault integration. 
Gradual rollout of cryptographic updates across projects or environments further reduces operational risk during migrations.  
In summary, GitLab CI/CD realizes cryptographic agility through pipeline extensibility, integration of internal services (NGINX, GitLab Shell, artifact signing), and external tool orchestration (cf.~GitLab~documentation\footnote{\url{https://docs.gitlab.com/development/architecture/}}~\footnote{\url{https://docs.gitlab.com/omnibus/settings/ssl/}}~\footnote{\url{https://gitlab.com/gitlab-org/gitlab}}).

\textbf{Constraints of cryptographic agility.} 
Agility in GitLab CI/CD is limited by the diversity of the environments it orchestrates. 
Runners, container images, and deployment targets must all support the chosen algorithms and parameters; otherwise, inconsistencies and failures occur. 
Because GitLab itself offers little native cryptographic functionality, support depends on external components whose correctness and timeliness must be assured separately. 
Coordinating upgrades across diverse teams and environments is complex, and GitLab provides only limited means to enforce uniform cryptographic policies across all pipelines.  
In summary, the main constraints on cryptographic agility in GitLab CI/CD are
dependencies on external components, the complexity of diverse environments, and limited native policy enforcement.

\textbf{Interoperability.} 
GitLab CI/CD influences cryptographic interoperability by orchestrating how algorithms and protocols are applied across software delivery pipelines and integrated services. 
TLS interoperability depends on the NGINX proxy configuration, while SSH compatibility is tied to GitLab Shell. 
Artifact verification requires alignment between signing tools (e.g.,~GPG, Cosign) and downstream verification processes. 
Registry interactions rely on consistent TLS and token handling across environments. 
By coordinating these internal and external components, GitLab extends interoperability across a wide range of tools and platforms. 
At the same time, differences in cryptographic support between components can lead to fragmentation, requiring careful coordination and phased migration. 
Thus, GitLab CI/CD exemplifies how interoperability in complex application environments is both enabled and constrained by the extent of cryptographic agility.

\subsection{Cross-Domain Dependencies}%
\label{sec:xdom_deps}
Across the three exemplars, a common pattern emerges: cryptographic agility at higher layers depends on changeability at lower ones, where OpenSSL provides the algorithmic foundation, NGINX exposes it through configuration, and GitLab CI/CD coordinates it across multiple integrated services.
This illustrates why cryptographic migrations require coordinated effort across
layers rather than isolated algorithm replacements.
In practice, practitioners should target the lowest layer that provides a stable abstraction (e.g.,~the OpenSSL provider interface) and enforce cryptographic policy as centrally as each layer permits.

The software domain highlights those cross-layer interactions, e.g., between the
\emph{library} and \emph{application} layer, especially well:
open-source software components expose the most direct and frequent mechanisms for cryptographic change, such as algorithm substitution and library updates, and are therefore most accessible for empirical analysis.
In contrast, the hardware and organizational domains evolve on longer timescales and
with limited public transparency. This constrains empirical observation and motivates focusing on software as the most analyzable layer.

Many real-world migration problems do not just need coordinated efforts across layers
but across domains. One example spanning all domains is software that requires
dedicated cryptographic hardware which might be, in turn, subject to supply chain
regulations.
Another concrete example is the interconnection between
the algorithm layer and the application layer for Partially Homomorphic
Encryption~(PHE). As their homomorphic properties are usually baked into their
applications, creating an abstraction layer like in OpenSSL~(cf.
Section~\ref{sec:ca_in_practice:openssl}) is significantly harder~\cite{phe_deps}.
Lightweight libraries such as LightPHE or TenSEAL do provide such an abstraction, exposing several partially homomorphic algorithms behind a unified interface, but swapping the underlying algorithm can shift the mathematical structure (e.g.,~from an additively to a multiplicatively homomorphic group), so the stable interface preserves the API surface but not the functional guarantees callers depend on.
Switching those applications to PQC-ready Fully Homomorphic Encryption~(FHE) schemes,
on the other hand, does not only change the whole software stack but also imposes
severe performance restrictions~\cite{he_pqc}. Consequently, these cross-domain
dependency-intensive instances are both harder to migrate and harder to make
cryptographically agile. However, cryptographic agility is especially important there,
as an ad-hoc migration is impeded by their inherent complexity.

The combination of the hardware and organizational domains being more opaque
and the complexity of domain-spanning dependencies makes analyzing those
cases more challenging and pushes it out of this paper's scope. Nevertheless, they
often represent important parts of our infrastructure and applying the crypto-agility
research in such large case studies is a crucial trajectory of future research.

\section{Discussion}%
\label{sec:discussion}
In this section, we build on the results of Section~\ref{sec:results} and their practical manifestations in Section~\ref{sec:ca_in_practice} in order to provide a broader discussion of cryptographic agility.
We highlight the inherent complexity of realizing it across contexts~(cf.~Section~\ref{sec:discussion:complexity_of_ca}), examine the challenges of assessing its extent~(cf.~Section~\ref{sec:discussion:assessment_of_cryptographic_agility}), and clarify its conceptual distinction from post-quantum cryptographic migration~(cf.~Section~\ref{sec:discussion:ca_vs_pqcmigration}).

\subsection{The Complexity of Agility and Interoperability}%
\label{sec:discussion:complexity_of_ca}
Introducing agility means introducing additional measures to the
respective entities. These include, e.g.,
flexible interfaces in the software domain or negotiating supplier contracts
containing options to change the procured hardware. At the same time,
some of these measures also help with other requirements. For example,
interfaces are also part of clean software engineering principles, and flexible contracts
are part of risk management. Thus, cryptographic agility comes with additional
work overhead and by that increases the complexity of the related projects but can
also reduce the complexity by limiting the amount of possible
(untracked) dependencies.

Striving for maximal interoperability across all contexts, on the other hand, can generate significant overhead. 
Large pools of supported algorithms and modes, combined with backward compatibility requirements, expand the codebase, add abstraction layers, and complicate decision logic. 
This complexity raises costs for implementation, testing, and maintenance~\citer{Ott19ResearchChallenges,Heftrig22PosterDNSSEC}. 
Furthermore, it can reduce security by enlarging the attack surface, prolonging the use of deprecated algorithms, and creating opportunities for misconfiguration.

Similarly, striving for high levels of cryptographic agility comes with engineering costs.
For example, provider architectures such as the one in OpenSSL require all consuming applications to use a stable abstraction layer consistently.
Legacy code that bypasses this abstraction must be rewritten to benefit from it.
Beyond implementation effort, migration periods in which old and new mechanisms coexist introduce security risks.
If negotiation and fallback behavior are not carefully controlled, such periods can increase the risk of downgrade attacks~\cite{downgrade, TLS_version_downgrade}.
Changeability is further bounded when the affected entity carries functional properties that must be preserved.
Partially homomorphic schemes illustrate this: a change must trade off cryptographic migration, performance overhead, and the functional guarantees the surrounding system depends on~(cf.~Section~\ref{sec:xdom_deps}).

Effective engineering of cryptographic systems therefore requires prioritization.
Insufficient agility leaves systems unable to respond to disruptive change, while
striving for maximal agility increases the implementation workload. As not every system will display perfect
agility, we will always need cryptographic interoperability to some extent but have
to keep in mind its inherent cost in complexity.
Consequently, organizations and system designers must decide whether the primary need is to ensure
resilience against emerging threats or to maintain compatibility with diverse peers,
and must implement agility or interoperability mechanisms accordingly.
Instead of generic solutions, the concepts must be realized in ways that align with the domain-specific requirements.
\\~\\
\begin{tcolorbox}[colback=gray!10,
	colframe=gray!50,
	title=\textbf{Consideration: Complexity vs. Agility and Interoperability},
	before=\nolinebreak,
	coltitle=black]
  \label{box:consideration_complexity_of_ca}
	Cryptographic agility comes at a cost of additional work but can even decrease
	complexity. High levels of cryptographic interoperability can, however, without
	clear focus, lead to excessive complexity and reduce security.
	Thus, it is essential to distinguish between \emph{cryptographic agility} and
	\emph{cryptographic interoperability} and to pursue targeted, context-specific
	solutions.
\end{tcolorbox}

\subsection{Assessing Cryptographic Agility}
\label{sec:discussion:assessment_of_cryptographic_agility}
In order to be able to make statements about the extent of the realized cryptographic
agility in a specific context (e.g.,~a device or software library), the question arises as to what extent cryptographic agility can be measured.  
Several approaches have been proposed, such as the Crypto-Agility Maturity Model (CAMM)~\citer{Hohm23MaturityModelCA}, which defines maturity levels with associated requirements, or the broader property sets introduced by Mehrez and Omri~\citer{Mehrez18CAProperties}. 
These models provide useful orientation but should not be mistaken for universally applicable metrics.  

A universal measurement is hardly feasible because cryptographic agility spans multiple dimensions. 
Depending on the context in which agility manifests, different cryptographic entities are affected and must meet different quality attributes. 
In a library, agility concerns the ability to exchange algorithm implementations efficiently, whereas at the organizational layer it relates to updating policies or suppliers with minimal disruption. 
These differences illustrate that cryptographic agility is not uniformly defined across contexts and that the associated quality attributes, such as efficiency, continuity, or flexibility, vary accordingly. 
Our layer model captures this heterogeneity by distinguishing between conceptual, software, hardware, and organizational domains and by specifying the contexts within them. 
It thus highlights that assessments cannot be generalized but must be tailored to the context in which agility is realized.
Our practical demonstration in Section~\ref{sec:ca_in_practice} covers the software domain only, while the hardware and organizational domains remain conceptual and require further empirical validation.

Against this background, it is more useful to view cryptographic agility as a spectrum that should be assessed relative to clearly defined requirements within a given context rather than by applying universal scales such as fixed maturity levels or percentage values.
In practice, this starts with clearly scoping the context and identifying which cryptographic entities can be changed and how.
OpenSSL illustrates this well: algorithms are exchanged through the provider architecture, and applications interact with them through the EVP API without needing to be recompiled.
Security policies such as disabling weak algorithms can be configured centrally for a single application via its OpenSSL configuration, though not across applications.
Practitioners can then evaluate concrete requirements such as whether algorithms can be replaced without modifying application code, whether cryptographic policies can be enforced centrally, and how much effort a migration to a new algorithm requires.
Additional requirements from models such as CAMM~\citer{Hohm23MaturityModelCA} or Mehrez and Omri~\citer{Mehrez18CAProperties} can then be applied to structure a more comprehensive assessment.  
\\~\\
\begin{tcolorbox}[colback=gray!10,
	colframe=gray!50,
	title=\textbf{Consideration: The Assessment of Cryptographic Agility},
	before=\nolinebreak,
	coltitle=black]
  \label{box:consideration_assessment_of_ca}
	\emph{Cryptographic agility} cannot be universally quantified, since the relevant entities and quality attributes vary across contexts. 
	Assessments should build on established models, but must always be tailored to the specific context in which agility is realized.
\end{tcolorbox}

\subsection{Distinguishing Cryptographic Agility from PQC Migration}
\label{sec:discussion:ca_vs_pqcmigration}
A frequent point of discussion is the relationship between cryptographic agility and post-quantum cryptographic (PQC) migration. 
Although often mentioned together, PQC migration and cryptographic agility are conceptually distinct. 
Cryptographic agility is a property of a given context, describing its capacity to add, enable, disable, and remove cryptographic entities in a sustainable and repeatable way. 
It describes how well the structure of systems, infrastructures, or organizations
accommodates change, regardless of whether this change is incremental or disruptive.
By contrast, PQC migration denotes a singular event: a targeted transition from
classical to post-quantum algorithms, typically triggered by external drivers such as
the emerging quantum threat, a regulatory mandate, or the completion of standardization milestones, as discussed by Näther et al.~\cite{naether2024PQCMigration}.
Migration, therefore, does not constitute a property of the system itself, but rather
a concrete process executed at a specific point in time. Considering our previously
introduced modes~(cf.~Section~\ref{sec:results:variety_of_definitions}), PQC
migration could be depicted as an objective.

Despite this difference, the two are tightly connected. 
The success and efficiency of PQC migration directly depends on the degree of agility present in the relevant contexts. 
In the software domain, modular libraries and flexible APIs determine whether algorithms can be substituted or extended without rewriting applications. 
In the hardware domain, devices such as HSMs or TPMs may either accelerate or constrain migration, depending on whether they can be upgraded to support new primitives. 
In the organizational domain, clear governance, procurement rules, and process structures provide the necessary framework to enforce deprecation of vulnerable algorithms and adoption of standardized PQC schemes.  

Moreover, PQC migration often requires an intentional reduction of interoperability.
To eliminate downgrade risks and meet standardization requirements, insecure legacy schemes must be disabled or removed, thereby narrowing the range of systems with which secure communication remains possible. 
Here agility provides the decisive factor: highly agile systems can disable or remove obsolete algorithms in a controlled and efficient way, while less agile ones risk prolonged coexistence of old and new cryptography, increasing exposure to downgrade attacks. 
Thus, cryptographic agility facilitates timely migration, whereas interoperability
mitigates the disruptions caused by entities whose lack of agility prevents swift
migration.
\\~\\
\begin{tcolorbox}[colback=gray!10,
	colframe=gray!50,
	title=\textbf{Consideration: Agility as Enabler of PQC Migration},
	before=\nolinebreak,
	coltitle=black]
  \label{box:consideration_ca_vs_pqc-migration}
  \emph{Cryptographic Agility} enables continuous change across contexts, whereas migration represents a one-time security objective.
  Without agility, migration becomes costly, fragmented, and error-prone; with agility, it can be executed in a structured, secure, and efficient manner. 
\end{tcolorbox}

\section{Threats to Validity}%
\label{sec:threats_to_validity}
In this section, we discuss potential threats to the validity of our study and how we mitigated them. We structure this discussion based on the widely recognized framework by Ampatzoglou et al.~\cite{Ampatzoglou20ThreatsToValidityGuidelines}, which categorizes threats into study selection validity, data validity, and research validity.

\subsection{Study Selection Validity}%
\label{sec:threats:study_selection_validity}
Study selection validity concerns the process of identifying and including
relevant studies in our analysis. In order to find the most relevant studies for
our research questions we constructed dedicated search strings for Google Scholar, the Google search engine, the ACM Digital Library, and IEEE Xplore. We used these four search engines for discovering
sources in several digital libraries to mitigate the resource bias. To moderate
risks associated with bias, we ensured an almost balanced number of authors from
academia ($n=29$), industry ($n=31$), and standardization bodies ($n=8$) as well as gray ($n=24$) and
peer-reviewed ($n=24$) literature. This is illustrated in
Table~\ref{t:descriptive_stats}. Additionally, we rigorously applied a set of
inclusion and exclusion criteria, ensuring that studies irrelevant to our research
were excluded~(cf. Table~\ref{tab:selection_criteria}).

We acknowledge, however, that we focused only on studies published in English.
While this may introduce a language bias, our preliminary search identified a
limited number of relevant studies in other languages, which we believe does not
significantly affect the validity of our results.
Furthermore, our use of Google Scholar and Google Search introduces a ranking bias, as both engines prioritize results based on popularity and citation counts rather than relevance to our research questions.
As a result, some sources may be captured unevenly, including preprints such as those on arXiv, particularly in a fast-moving area where terminology is still evolving.

\subsection{Data Validity}
Data validity relates to the accuracy and reliability of the data extracted from
the selected studies. Given the relatively small number of papers included in
our analysis ($n=48$), we consider the set to be highly relevant to our research
questions and representative of the current state of the field. Furthermore, we
employed multiple reviewers during data extraction to
minimize subjective bias and random errors. Each reviewer performed independent
checks, and we conducted cross-validation to confirm consistency in data
interpretation.

Although we did not perform a formal quality assessment of the included studies,
our strict inclusion criteria ensure that only papers with clear methodologies
and substantial contributions to the field are considered. This approach helps
exclude lower-quality publications regardless of their format.
Including gray literature nonetheless broadens the evidence base and reduces publication bias.
Because these sources still differ in rigor, some unevenness in coverage may remain, which adds to the ranking bias noted in Section~\ref{sec:threats:study_selection_validity}.

\subsection{Research Validity}
Ampatzoglou et al.~\cite{Ampatzoglou20ThreatsToValidityGuidelines} mention that
the lack of repeatability, the lack of coverage of research questions, the lack
of comparable related work and the lack of generalizability threaten the
validity of studies.

As we documented the exact search strings used for all four search
engines, our search strategy can be reproduced. We also provide a replication
package for the data extraction so that independent researchers are able to
reproduce the whole procedure.
Given the fast-moving nature of PQC-related standardization and terminology, we note that the present review represents a snapshot of the literature at the time of search.
The replication package~\cite{replication_package} can serve as a starting point for follow-up search passes as the field evolves. Furthermore, we ensured the coverage of our
research questions by the precise formulation following the PRISMA~2020
statement~\cite{Prisma2020}. Within Section~\ref{sec:results}, which lists our
selected papers (see~Table~\ref{tab:category_coverage}), we made sure to include
the most relevant articles in the context of cryptographic agility. We did not
find a similarly systematic approach in the literature. We believe that our study
is sufficiently generalizable, as it encompasses a broad spectrum of academic
and gray literature from various venues, providing a comprehensive overview of
the topic at hand.

\subsection{Definition Validity}
Within RQ1 we establish six categories in order to map every aspect of the reviewed
definitions to one of these categories.

As we have only used the reviewed literature for this, we cannot make any
statements about the general validity and complete coverage of our categories.
The categories so far fully cover all cryptographic agility definitions we have
encountered in our work, as demonstrated in Table~\ref{tab:category_coverage}.
All of the aspects in these definitions can be mapped to one of the categories
we defined.

Given the wide range of definitions, certain elements may have been incorporated
into the literature-based definition of cryptographic agility in ways that could
benefit from refinement. For example, in the context of data synthesis,
interoperability was initially considered part of business continuity, which
does not align with the current understanding of the concept. We addressed this
challenge with our fourth research question, aiming to clarify and distinguish
existing terminology and to develop a new taxonomy that better aligns with the
evolving needs in this field of research.

In conclusion we employ a robust methodology and adhere to well-established
frameworks and practices. We believe these efforts have minimized the impact of
potential biases and errors in our study, ensuring the reliability of our
conclusions.

\section{Conclusion}%
\label{sec:conclusion}
This paper provides the first systematic study of cryptographic agility. 
We synthesized prior attempts at definition, refined them, and established a conceptual foundation that clarifies its scope and relation to neighboring terms.

To answer RQ1, we analyzed existing definitions and mapped their aspects into six categories: \emph{context}, \emph{mode}, \emph{desired capabilities}, \emph{quality attributes}, \emph{cryptographic entities}, and \emph{drivers}. 
To answer RQ2, we synthesized these results into a literature-based definition that unifies the most significant characteristics across prior work, providing the first systematic definition derived entirely from the existing body of knowledge. 
To answer RQ3, we extended this foundation conceptually and derived a refined, context-independent definition that isolates agility as the property of changeability of cryptographic entities. 
To answer RQ4, we introduced a layer model that organizes contexts into four
overarching domains---conceptual, software, hardware, and organization---and clarified the relation of agility to the related concepts of versatility and interoperability.

We demonstrated the applicability of our results through three real-world exemplars
from the software domain: OpenSSL, NGINX, and GitLab CI/CD\@.
These cases showed how agility is realized in practice, which constraints limit it,
and how it interacts with interoperability.
The exemplars validate the framework in the software domain only. The other domains remain conceptual.
Future work should extend the empirical validation beyond software to hardware and organizational entities, especially case studies with strong cross-domain dependencies, and investigate context-specific assessment approaches for cryptographic agility.

In our discussion, we highlighted broader considerations that extend beyond individual cases. 
We showed how cryptographic agility is both necessary and complex, how its assessment depends on contextual factors, and how it is closely tied to migration scenarios such as post-quantum cryptography. 

Taken together, our results establish a clear conceptual foundation for cryptographic agility. 
They provide terminology, structure, and examples that enable consistent reasoning across contexts and domains. 
We expect this work to guide future research and practical efforts in building systems that can accommodate cryptographic change securely, efficiently, and sustainably.

\newpage

\bibliographystyler{IEEEtran}
\bibliographyr{bibliography}
\bibliographystyle{IEEEtran}
\bibliography{bibliography}

~\newpage

\unless\ifanonymous
  \begin{IEEEbiography}[{\includegraphics[width=1in,height=1.25in,clip,keepaspectratio]{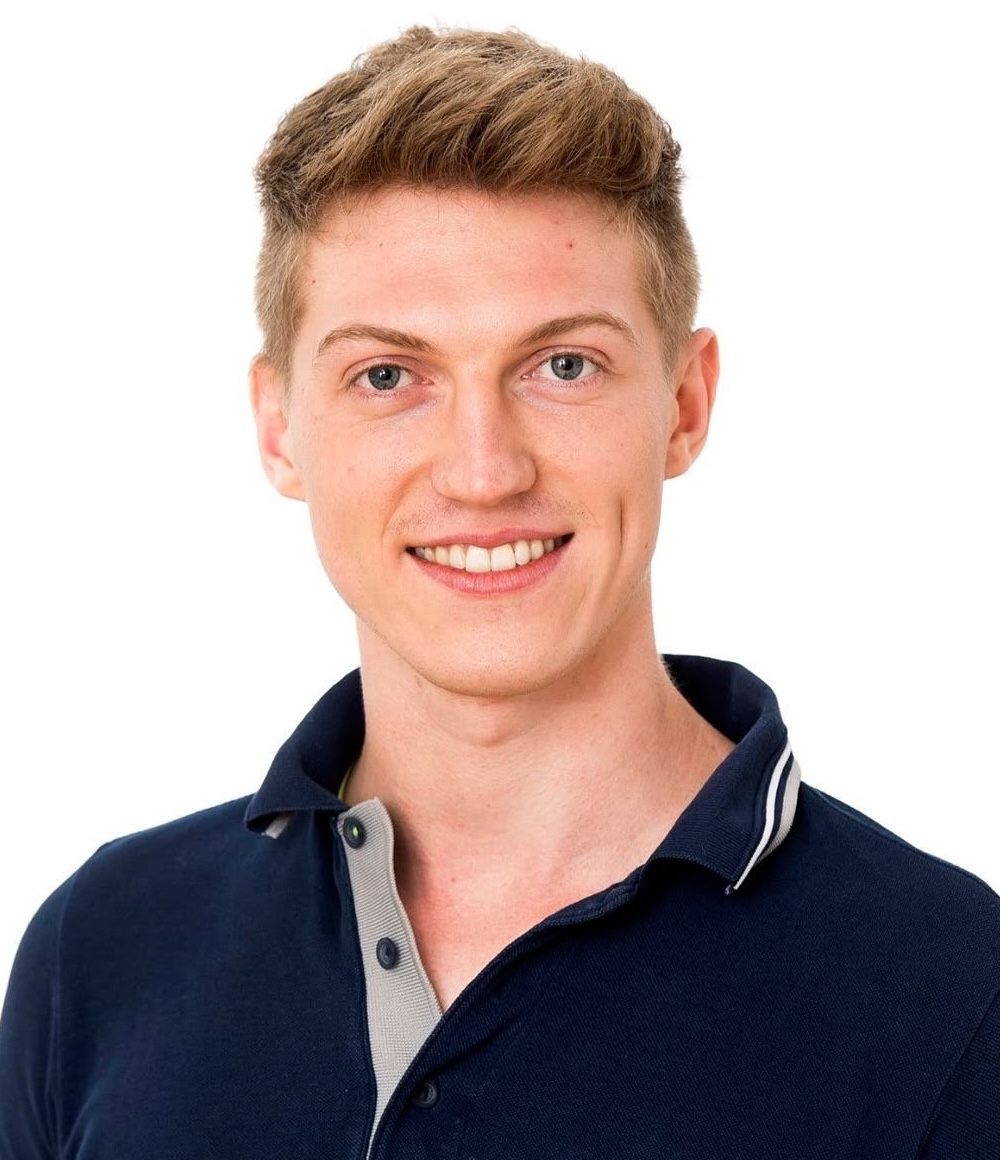}}]
  {Christian Näther} received his Master's degree in Computer Science from the University of Augsburg, Germany.
  As a former software engineer, he has gained several years of experience in the fields of software development, distributed systems as well as information security. 
  He is currently a security researcher at XITASO and part of the research project AMiQuaSy, which focuses on the migration of systems towards post-quantum cryptography.
  Christian Näther is an active board member of XITASO’s information security community, where he promotes both internal and customer-specific security. 
\end{IEEEbiography}
  
\begin{IEEEbiography}[{\includegraphics[width=1in,height=1.25in,clip,keepaspectratio]{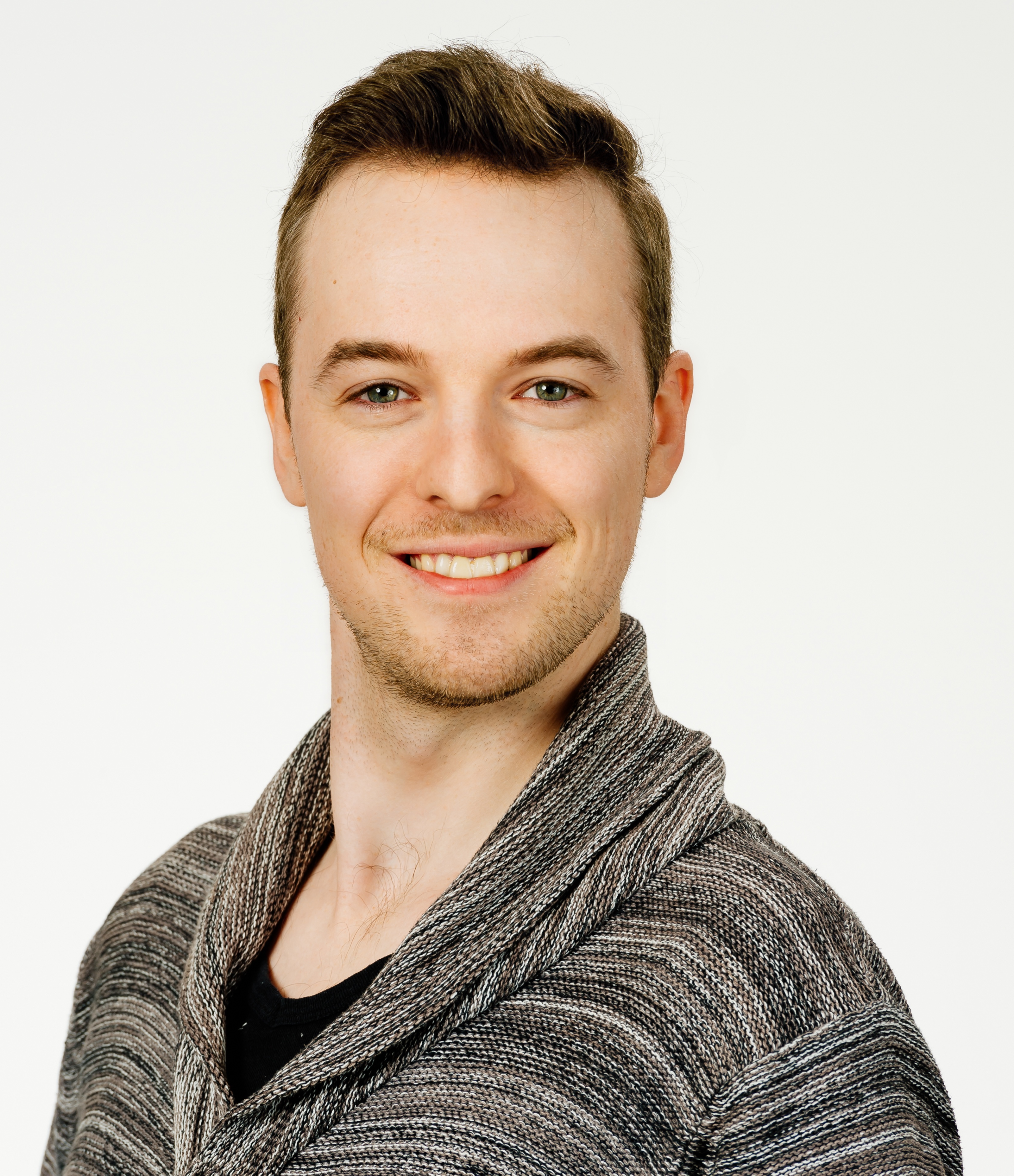}}]
{Daniel Herzinger} holds a Master of Science in Computer Science.
He started working on post-quantum cryptography (PQC) and its adoption into real-world scenarios during his Master's thesis in 2020, together with the German IT security company genua, where he started his IT security journey in 2015. After finishing his studies, he first dived into supporting customers with securing their IT and OT infrastructures at genua's consulting department for two years. He then indulged in his passion for PQC by joining the research group in 2023, focusing on the transition of complex systems and infrastructures to quantum resistance.
\end{IEEEbiography}

\begin{IEEEbiography}[{\includegraphics[width=1in,height=1.25in,clip,keepaspectratio]{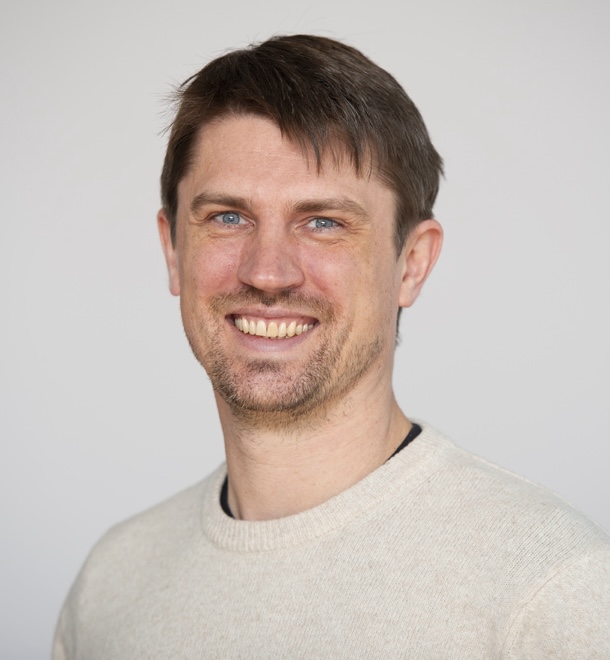}}]
{Jan-Philipp Steghöfer} (Member, IEEE) received the Ph.D. degree in Computer Science from the University of Augsburg, Germany.
As a former professor at the University of Gothenburg he has an extensive and distinguished academic background with a strong passion for research and innovation.
His research interests include security assurance, security of AI systems, agile development in regulated domains, and model-driven engineering.
He is currently Head of Research at XITASO as well as an active member of the software engineering community where he is part of the organisation committee for events like the IEEE International Requirements Engineering Conference (RE). 
He also reviews for International Conference on Software Engineering (ICSE), IEEE Transactions on Software Engineering (TSE), Journal of Systems and Software (JSS), Empirical Software Engineering and Measurement (ESEM), and many other venues.
\end{IEEEbiography}

~\newpage

\begin{IEEEbiography}[{\includegraphics[width=1in,height=1.25in,clip,keepaspectratio]{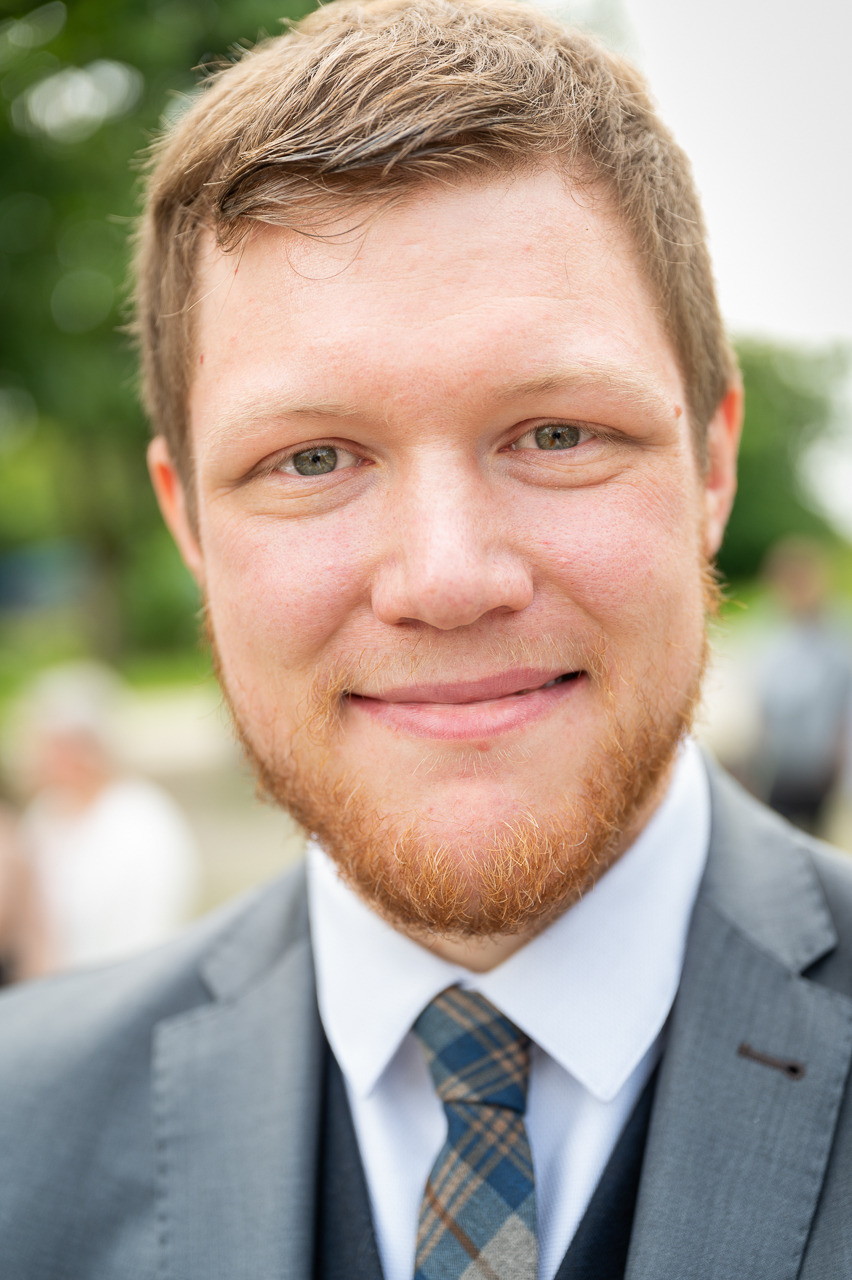}}]
  {Stefan-Lukas Gazdag} holds a Master of Science in Computer Science. He has focused on post-quantum cryptography (PQC) and future-proof security mechanisms since 2013, overcoming obstacles in practical use. Being part of the German IT security company genua’s research group he currently works on the fourth publicly funded PQC research project and other collaborations, enabling the transition to quantum-safe networks. He was e.g. involved in the publication of the first explicit PQC RFC, RFC 8391 describing the signature scheme XMSS, recommended e.g. by US NIST, NSA and German BSI.
  \end{IEEEbiography}

\begin{IEEEbiography}[{\includegraphics[width=1in,height=1.25in,clip,keepaspectratio]{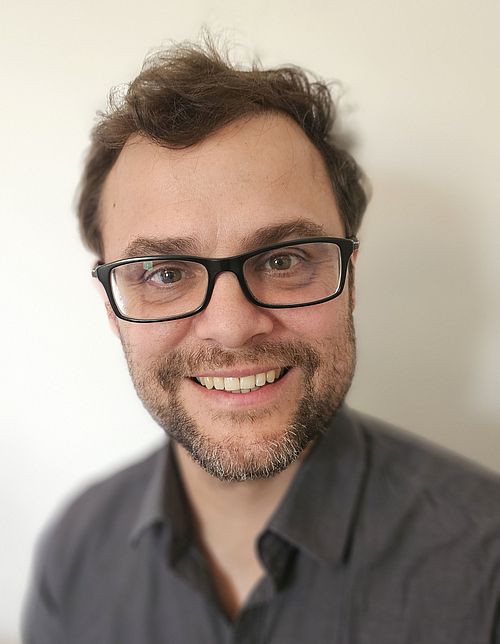}}]
{Eduard Hirsch} holds a Master of Science degree in Computer Science from
the Technical University of Vienna. Over the years he accumulated extensive experience in industry, contributing to various projects in the software and technology sector. In addition to his professional work, he has been teaching and doing research at the Salzburg University of Applied Sciences, where he was contributing to national (Austria) and international research projects (e.g. H2020, Interreg). His research interests span software engineering, industrial automation, big data engineering as well as security. He is currently a researcher at the
Technical University Amberg-Weiden, where he focuses on advancing both practical and theoretical aspects of PQC migration but also teaches in the area of security.
\end{IEEEbiography}

\begin{IEEEbiography}[{\includegraphics[width=1in,height=1.25in,clip,keepaspectratio]{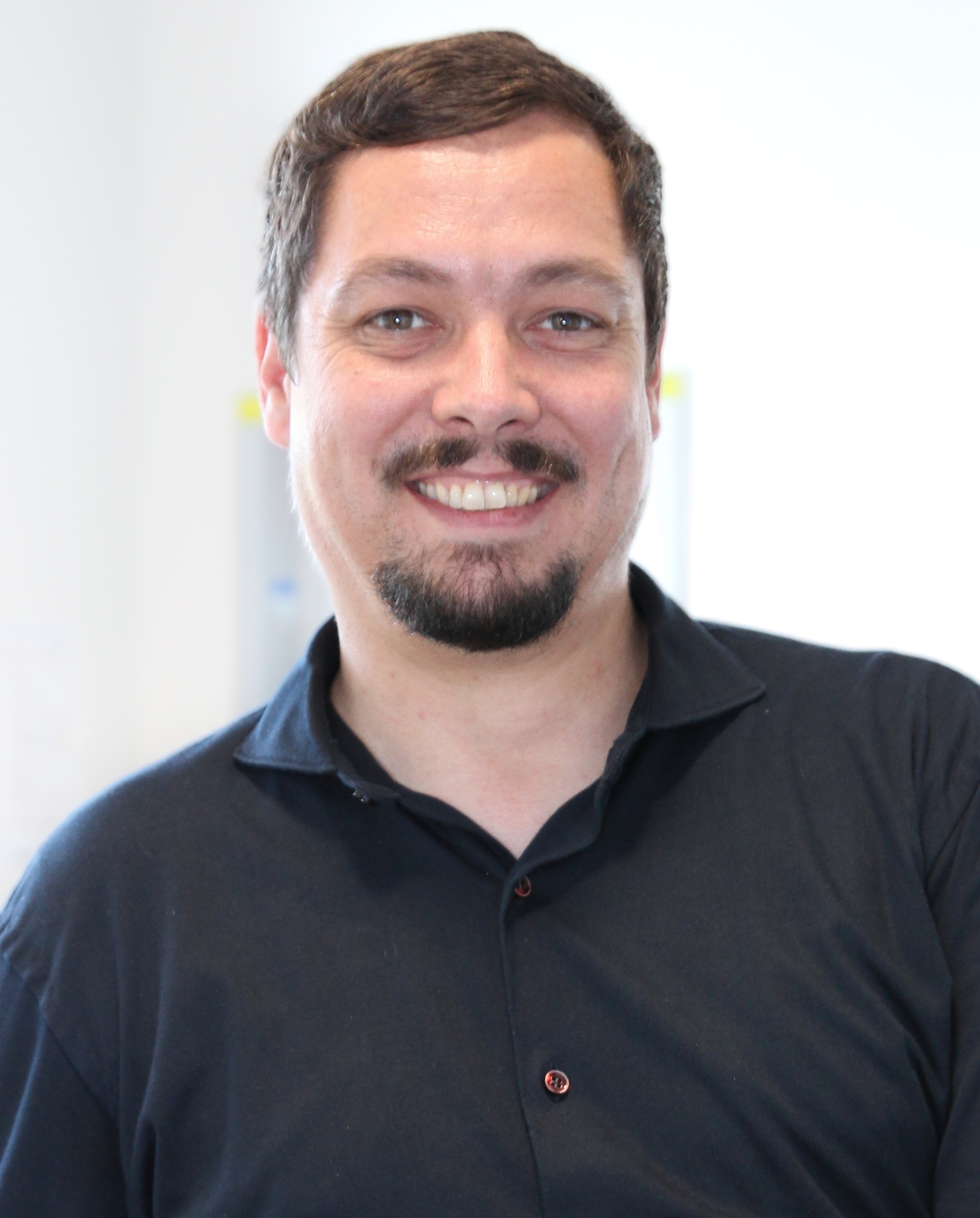}}]
{Daniel Loebenberger} received his doctorate in cryptography from the University of Bonn in 2012, where he conducted research and taught until the end of 2015. From 2016 to 2019, he worked as an IT security expert with a focus on cryptography at genua GmbH, a subsidiary of the Bundesdruckerei GmbH, the German Federal Printing Office, on various topics of professional high-security components. Since January 2019, Daniel Loebenberger has been appointed Professor of Cybersecurity at the Technical University Amberg-Weiden and also heads the department ``Secure Infrastructure'' at Fraunhofer AISEC's Weiden site. In the lab there, topics of applied post-quantum cryptography and quantum-safe infrastructures are addressed in research and teaching.
\end{IEEEbiography}
\fi

\EOD

\end{document}